\documentclass[11pt,a4paper]{article}
\pdfoutput=1

\usepackage{amsmath,amsfonts,amsbsy,amssymb,enumerate,array}
\usepackage{graphicx,latexsym,verbatim,psfrag,dsfont}
\usepackage[usenames]{color}
\usepackage[english]{babel} 
\usepackage{fancybox}
\usepackage{slashed}

\usepackage{bm} 
\usepackage{booktabs} 
\usepackage{chngpage} 

\usepackage{psfrag}
\usepackage{enumitem}
\setlist{itemsep=0pt}

\usepackage{fancyhdr}
\usepackage{datetime}

\usepackage[nosort]{cite}   
\usepackage{mciteplus}

\usepackage[colorlinks=true,      linkcolor=blue,      urlcolor=blue,      
filecolor=blue,      citecolor=blue,       pdfstartview=FitH,     
pdfpagemode=UseNone,      bookmarksopen=true, linktocpage=true]{hyperref}  
\usepackage[all]{hypcap}     

\begin{document}
	
	\numberwithin{equation}{section}
	

	\newcommand{\be}{\begin{equation}}
	\newcommand{\ee}{\end{equation}}
	\newcommand{\bea}{\begin{eqnarray}\displaystyle}
	\newcommand{\eea}{\end{eqnarray}}
	
	\def\eq#1{(\ref{#1})}
	\newcommand{\fig}{Fig.\;\ref}
	
	\newcommand{\captionfonts}{\small}
	
	\makeatletter  
	\long\def\@makecaption#1#2{%
		\vskip\abovecaptionskip
		\sbox\@tempboxa{{\captionfonts #1: #2}}%
		\ifdim \wd\@tempboxa >\hsize
		{\captionfonts #1: #2\par}
		\else
		\hbox to\hsize{\hfil\box\@tempboxa\hfil}%
		\fi
		\vskip\belowcaptionskip}
	\makeatother   
	
	
	\def\a{\alpha}  \def\b{\beta}   \def\c{\chi}
	\def\g{\gamma}  \def\G{\Gamma}  \def\e{\epsilon}
	\def\vep{\varepsilon}   \def\tvep{\widetilde{\varepsilon}}
	\def\f{\phi}    \def\F{\Phi}  \def\fb{{\ov \phi}}
	\def\vf{\varphi}  \def\m{\mu}  \def\mub{\ov \mu}
	\def\n{\nu}  \def\nub{\ov \nu}  \def\o{\omega}
	\def\O{\Omega}  \def\r{\rho}  \def\k{\kappa}
	\def\kab{\ov \kappa}  \def\s{\sigma}
	\def\t{\tau}  \def\th{\theta}  \def\sb{\ov\sigma}  \def\S{\Sigma}
	\def\l{\lambda}  \def\L{\Lambda}  \def\p{\psi}
	
	\newcommand{\gt}{\tilde{\gamma}}
	
	
	\def\cA{{\cal A}} \def\cB{{\cal B}} \def\cC{{\cal C}}
	\def\cD{{\cal D}} \def\cE{{\cal E}} \def\cF{{\cal F}}
	\def\cG{{\cal G}} \def\cH{{\cal H}} \def\cI{{\cal I}}
	\def\cJ{{\cal J}} \def\cK{{\cal K}} \def\cL{{\cal L}}
	\def\cM{{\cal M}} \def\cN{{\cal N}} \def\cO{{\cal O}}
	\def\cP{{\cal P}} \def\cQ{{\cal Q}} \def\cR{{\cal R}}
	\def\cS{{\cal S}} \def\cT{{\cal T}} \def\cU{{\cal U}}
	\def\cV{{\cal V}} \def\cW{{\cal W}} \def\cX{{\cal X}}
	\def\cY{{\cal Y}} \def\cZ{{\cal Z}}
	
	\def\mC{\mathbb{C}} \def\mP{\mathbb{P}}
	\def\mR{\mathbb{R}} \def\mZ{\mathbb{Z}}
	\def\mT{\mathbb{T}} \def\mN{\mathbb{N}}
	\def\mH{\mathbb{H}} \def\mX{\mathbb{X}}
	
	\def\one{{\hbox{\kern+.5mm 1\kern-.8mm l}}}
	
	
	\newcommand{\bra}[1]{{\langle {#1} |\,}}
	\newcommand{\ket}[1]{{\,| {#1} \rangle}}
	\newcommand{\braket}[2]{\ensuremath{\langle #1 | #2 \rangle}}
	\def\corr#1{\left\langle \, #1 \, \right\rangle}
	\def\vac{|0\rangle}
	\newcommand{\pA}{{\text{\tiny A}}}
	\newcommand{\pB}{{\text{\tiny B}}}

	
	\def\d{ \partial }
	\newcommand{\floor}[1]{\left\lfloor #1 \right\rfloor}
	\newcommand{\ceil}[1]{\left\lceil #1 \right\rceil}
	\newcommand{\CR}{\nonumber \\*}
	\newcommand{\?}{\;\!}
	
	\mathchardef\mhyphen="2D
	
	\def\nn{\nonumber}
	
	
	\newcommand{\Scal}[1]{\Bigl ({#1} \Bigr )}
	\newcommand{\scal}[1]{\bigl ({#1} \bigr )}
	\def\ie{{\it i.e.}\ }
	\def\eg{{\it e.g.}\ }
	
	\definecolor{cardinal}{rgb}{0.6,0,0}
	\definecolor{darkgreen}{rgb}{0,0.4,0}
	\definecolor{darkblue}{rgb}{0, 0, 0.7}
	\newcommand{\Red}{\color{red}}
	\newcommand{\Blue}{\color{blue}}
	\newcommand{\DarkGreen}{\color{darkgreen}}
	\newcommand{\DG}{\color{darkgreen}}
	\newcommand{\Cardinal}{\color{cardinal}}
	\newcommand{\DarkBlue}{\color{darkblue}}

	\newcommand{\tblue}[1]{\scriptsize{\Blue #1}}
	
	
	\newcommand{\bV}{\overline{V}}
	\newcommand{\Vb}{\overline{V}}
	
	\newcommand{\GMR}{\rm GMR}
	
	\def\ax{\alpha}
	
	\newcommand{\ft}[2]{{\textstyle\frac{#1}{#2}}}

	\def\bh{\mathrm{\sst{BH}}}
	\def\btz{\mathrm{\sst{BTZ}}}
	
	\def\phibtz{\varphi}
	\def\lads{\ell_{\mathrm{\sst{AdS}}}}
	
	\newcommand{\adstwo}{AdS$_2$}
	\newcommand{\adsthree}{AdS$_3$}
	
	\newcommand{\adsthreetimessthree}{AdS$_3\times$S$^3$}

	\newcommand{\np}{\ensuremath{n_{\mathrm{\sst{P}}}}}
	\newcommand{\qp}{\ensuremath{Q_{\mathrm{P}}}}
	
	\newcommand{\lambdafour}{\lambda_{4}}
	\newcommand{\lambdahatfour}{\hat\lambda_{4}}
	
	\newcommand{\thetafour}{\Theta_{4}}
	\newcommand{\thetahatfour}{\medhat\Theta_{4}}
	\newcommand{\eu}{\mathrm{e}}
	
	\newcommand{\T}[3]{\ensuremath{ #1{}^{#2}_{\phantom{#2} \! #3}}} 
	
	\def\rgh{\ensuremath{r_{\mathrm{\sst{GH}}}}}
	\def\rhogh{\ensuremath{\rho_{\mathrm{\sst{GH}}}}}
	\def\ident{\ensuremath{\sim}}
	\def\yan{\ensuremath{\mathsf{y}}}
	

	\begin{flushright}
		CPHT-RR086.112017\\
		IPhT-T17/175\\
	\end{flushright}
	
	\vspace{16mm}
	
	\begin{center}
		{\LARGE \bf {Two Kissing Bolts}}
		
		\vspace{15mm}

		{\bf Guillaume Bossard$^{1,2}$, ~Stefanos Katmadas$^{3,4}$, ~David Turton$^5$}

		\vspace{15mm}


		\begin{small}
			
			{\it ${}^1$ Centre de Physique Th\'eorique, Ecole Polytechnique, CNRS, Universit\'e Paris-Saclay,\\
				91128 Palaiseau Cedex, France}
			
			\vspace{1pt}
			
			{\tt guillaume.bossard at cpht.polytechnique.fr}

			\vspace{10pt}

			{\it ${}^2$ Universit\`a di Roma Tor Vergata, Dipartimento di Fisica,\\
				Via della Ricerca Scientifica, I-00133 Rome, Italy}

			\vspace{10pt}
			
			{\it ${}^3$ Institut de Physique Th\'eorique, Universit\'e Paris Saclay, CEA, CNRS\\
				Orme des Merisiers,  F-91191 Gif sur Yvette, France}
			
			\vspace{10pt}
			
			{\it ${}^4$ Instituut voor Theoretische Fysica, Katholieke Universiteit Leuven\\
				Celestijnenlaan 200D, B-3001 Leuven, Belgium}
			
			\vspace{1pt}

			{\tt stefanos.katmadas at kuleuven.be}
			
			\vspace{10pt}
			
			{\it ${}^5$  Mathematical Sciences and STAG Research Centre, University of Southampton,\\ Highfield,
				Southampton SO17 1BJ, United Kingdom}
			
			\vspace{1pt}
			
			{\tt  d.j.turton at soton.ac.uk}
			
		\end{small}

		\vspace{18mm}

		\baselineskip=15pt
		\parskip=0pt

		\textsc{Abstract}

		\begin{adjustwidth}{15mm}{15mm} 
			\vspace{4mm}
			\noindent
			The study of non-supersymmetric black hole microstates offers the potential to resolve the black hole information paradox. 
			A system of equations was recently obtained that enables the systematic construction of non-supersymmetric smooth horizonless supergravity solutions, that are candidates to describe microstates of non-extremal black holes.
			Within this system we construct a family of six-dimensional supergravity solutions that feature two topologically-nontrivial three-cycles known as bolts. The two bolts touch at a single point and are supported by fluxes. 
			We find that the fluxes on the two three-cycles can be either aligned or anti-aligned, and exhibit examples of both. 
			We present several examples of smooth solutions, including near-extremal solutions that have an approximate AdS$_3$ region, and far-from extremal solutions that have arbitrarily small charge compared to their mass. 
		\end{adjustwidth}

	\end{center}

	\thispagestyle{empty}

	\newpage
	
	\baselineskip=15.2pt
	\parskip=2pt
	
	\setcounter{tocdepth}{2}
	\setcounter{page}{1}
	\tableofcontents
	
	\section{Introduction and Discussion}
	
	There has been significant renewed interest in the black hole information paradox~\cite{Hawking:1976ra,Mathur:2009hf}, and its implications for the physics of an observer falling into a black hole~\cite{Almheiri:2012rt,Mathur:2012jk,Almheiri:2013hfa,Mathur:2013gua}. 
	In String Theory, black hole entropy is understood as arising from an exponential degeneracy of internal microstates of strings and branes~\cite{Strominger:1996sh}. 
	This fact alone is not sufficient to resolve the information paradox.
	However, there are indications that the gravitational description of black hole microstates in String Theory may involve non-trivial physics at the horizon scale that, if sufficiently generic, could resolve the information paradox; for reviews, see~\cite{Mathur:2005zp,Bena:2007kg,Skenderis:2008qn,Balasubramanian:2008da,Mathur:2012zp,Bena:2013dka}.
	
	These indications are strongest for small two-charge supersymmetric black holes, for which there is a complete account of microstates~\cite{Lunin:2001jy,Lunin:2002iz,Rychkov:2005ji,Kanitscheider:2007wq}.
	However, it remains an open problem to determine whether such non-trivial structure at the horizon exists also for large black holes, and when supersymmetry is absent.
	
	Several families of microstates of large supersymmetric black holes admit descriptions as smooth horizonless solutions to supergravity that have the same mass and charges as the corresponding black hole, and that have a known description in the holographically-dual conformal field theory ~\cite{Giusto:2004id,Giusto:2004ip,Giusto:2012yz,Bena:2015bea,Bena:2016agb,Bena:2016ypk,Bena:2017geu}. Recent studies have uncovered interesting physics of probe particles on some of these backgrounds~\cite{Bena:2017upb,Tyukov:2017uig}, and of the stability properties of some supersymmetric solutions~\cite{Eperon:2016cdd,Marolf:2016nwu}. There are also large classes of supersymmetric solutions whose holographic description is not known; for a non-exhaustive sample see~\cite{Bena:2005va,Berglund:2005vb,Bena:2006kb,Bena:2017fvm,Avila:2017pwi}.
	
	However, for non-extremal black holes the situation is far less understood, as the task of constructing non-supersymmetric supergravity solutions is more difficult than in the supersymmetric case. For several years, the only known non-extremal black hole microstate solutions were the solution obtained via analytic continuation of the Cveti\v{c}-Youm black hole \cite{Cvetic:1996xz} by Jejjala, Madden, Ross and Titchener (JMaRT) \cite{Jejjala:2005yu}, as well as generalisations thereof~\cite{Giusto:2007tt,AlAlawi:2009qe,Banerjee:2014hza}. The JMaRT solution is asymptotically $\mathds{R}^{1,4}\times $S$^1$ with the same mass and charges as a five-dimensional three-charge non-extremal Cveti\v{c}-Youm black hole. The solution is smooth and horizonless; in its core the geometry caps off smoothly with a single topologically-nontrivial three-cycle that we refer to as a bolt. In an appropriate near-extremal regime, the local geometry near the cap becomes global AdS$_3 \times$S$^3$, with possible discrete identifications.
	
	In general, by a ``bolt'' we denote a locus of space diffeomorphic to the centre of $\mathds{R}^2$ times a compact surface (or a discrete quotient thereof), which in this case is a three-sphere. The terminology is a generalisation of that of~\cite{Gibbons:1979xm}.
	The bolt cycle of the JMaRT solution is supported by three-form flux that gives rise to the total electric charge of the solution at infinity. While this structure is analogous to the known bubbling microstate geometries for BPS black holes, the JMaRT solution can at best be viewed as a very atypical microstate, given its simple structure and the fact that both its angular momenta are always above the black hole regularity bound.

	To obtain more general families of solutions describing microstates of non-extremal black holes, one requires a systematic method.
	In recent work, a partially-solvable system of differential equations has been constructed, that enables the construction of much larger classes of supergravity solutions~\cite{Bossard:2014ola, Bena:2015drs, Bena:2016dbw}.
	This system consists of a sequence of second-order linear differential equations defined on a three-dimensional base metric, identified as the base metric of an auxiliary four-dimensional gravitational instanton. The first layer of the system, as formulated in \cite{Bena:2016dbw}, comprises the equations for this four-dimensional gravitational instanton and is the only non-linear layer; all following layers are linear. 
	The four-dimensional gravitational instanton is auxiliary in the sense that it does not appear geometrically as part of the six-dimensional metric of the solutions obtained by this method.
	
	Solutions to this system include the JMaRT solution; in this formulation, the first-layer data of the JMaRT solution correspond to a Kerr-NUT four-dimensional gravitational instanton.
	One can intuitively think of the Kerr-NUT two-sphere bolt as providing the seed for the JMaRT three-sphere bolt. 
	
	Starting from the same Kerr-NUT instanton, this system has enabled the construction of smooth solutions with at least one, and potentially an arbitrary number, of Gibbons--Hawking centres (nuts) in an axisymmetric configuration~\cite{Bena:2015drs,Bena:2016dbw}. These centres and the resulting topological cycles, usually referred to as bubbles, are of the same type appearing in BPS solutions, and indeed become BPS in the appropriate limit.
	The resulting solutions then carry an arbitrary number of these ``extremal'' bubbles together with a single bolt that is responsible for the additional energy above the BPS bound. 
	We distinguish bolts from extremal bubbles as follows: a bolt is a rigid three-cycle attached to a specific locus in space (the centre of $\mathds{R}^2$ mentioned above), while an extremal bubble is a cycle that is only determined by two specific points that we refer to as Gibbons--Hawking centres or ``nuts'',\footnote{In our solutions, we denote by a ``nut'' a point whose neighbourhood is diffeomorphic to a discrete quotient of $\mathds{R}^4 \times $S$^1$. A more detailed discussion is given in~\cite{Bena:2015drs}.} and is therefore not attached to a fixed surface a priori. Within this system, as formulated in \cite{Bena:2016dbw}, it appears to be a general feature that bolts arise if and only if they are already present in the gravitational instanton seeding the solution, whereas nuts appear at poles of the solutions to the linear equations defined over its three-dimensional base.

	Based on these constructions, one can extract a candidate picture of a class of microstate geometries of non-extremal black holes. This class consists of smooth horizonless solutions whose topological structure is identified as collections of rigid three-cycles (bolts) and Gibbons--Hawking centres (nuts), where the former are responsible for the energy above extremality. This is a direct generalisation of the multi-centre supersymmetric bubbling geometries, where only nuts arise. For example, the JMaRT solution admits a BPS limit to a two-nut supersymmetric solution, \ie with a single bubble, while the solutions described in \cite{Bena:2015drs, Bena:2016dbw} admit a BPS limit in which the single bolt becomes an additional non-rigid cycle attached to two nuts, among the other nuts.  
	
	To put this picture to the test, one would like to construct solutions with several nuts and bolts. Since in the system of~\cite{Bena:2016dbw} bolts arise from the first-layer data, this requires building solutions starting from four-dimensional gravitational instantons containing several bolts. This is a rather nontrivial task, because very little is known about such gravitational instantons with multiple bolts.
	
	However, there is a four-dimensional gravitational instanton solution that contains two Kerr-NUT bolts, touching at a point along their common axis of rotation~\cite{Chen:2011tc,Chen:2015vva}. This is a six parameter solution: the Kerr-NUT bolts carry three parameters each. The solution has three special points, and it reduces to a three-centre Gibbons--Hawking solution in an appropriate extremal limit. Unlike the multi-Taub-NUT instantons described in \cite{Gibbons:1979nf}, this solution does not exhibit the lines of conical singularities that are characteristic of unstable configurations.\footnote{Such singularities can be understood as compensating the gravitational forces arising in Minkowski signature. The instantons of \cite{Chen:2011tc,Chen:2015vva} cannot be analytically continued to Minkowski signature.} Therefore this solution is appropriate for use in the system of~\cite{Bena:2016dbw}.
	
	In this paper we construct a family of six-dimensional supergravity solutions based on the instanton of~\cite{Chen:2011tc,Chen:2015vva}. The resulting solutions feature two bolts touching at one point, and supported by three-form fluxes. Upon imposing the $\mathds{R}^{1,4}\times $S$^1$ asymptotics appropriate for a black hole, as well as smoothness and absence of closed timelike curves (CTCs), we find a family of smooth horizonless solutions with the same mass and charges as the non-extremal rotating three-charge black hole in five dimensions. Our family of solutions includes additional physical parameters describing the structure of the solution in its interior, in particular the fluxes and topology of the two independent three-cycles. These solutions have a natural limit to the JMaRT solution, in which one special point disappears and only a single bolt remains. 
	
	We analyse several interesting properties of our family of solutions. In particular, we find a set of near-extremal solutions that feature approximate AdS$_3$ throats.
	We also describe a sub-family of solutions that have arbitrarily small charge compared to their mass, and discuss their regularity. In both cases, as well as in our more general computer-aided scans of the solution space, we have only found solutions that have at most one of the two angular momenta within the regularity bound. In many examples the angular momentum that lies outside the bound, $J_\psi$, has a value very close to the bound, with a ratio $\tfrac{|J_\psi|-J_{\text{max}} }{J_{\text{max}} } \simeq 10^{-6}$ compared to the allowed maximum, $J_{\text{max\,}}$. This includes the example with an approximate AdS$_3$ throat exhibited in detail in Section \ref{sec:AdS3}.
	These properties are reminiscent of the situation early in the development of supersymmetric bubbling microstate geometries, which for a small number of centres also exhibit an angular momentum above the regularity bound. This obstacle can be overcome either by breaking isometries~\cite{Bena:2016ypk} or by adding more centres (see for example \cite{Bena:2017fvm,Avila:2017pwi} for recent constructions of such solutions). Since our two-bolt family of solutions is based on only three collinear centres, it is natural to expect that one cannot bring both angular momenta below the regularity bound; indeed the same situation arose in the three-centre solutions of \cite{Bena:2015drs}. Based on the success of supersymmetric solutions when isometries are broken\footnote{Note also the recent work on perturbative constructions of non-supersymmetric superstrata~\cite{Bombini:2017got}.} 
	or more centres are added, looking to the future we believe that there is cause for optimism on this point. 
	
	The paper is organised as follows. 
	In Section \ref{sec:sugra-sol} we present both the partially-solvable system of \cite{Bena:2016dbw} as well as the gravitational instanton of \cite{Chen:2011tc,Chen:2015vva}, before describing the construction of our family of six-dimensional supergravity solutions. 
	We proceed to impose smoothness on the family of solutions in Section \ref{sec:regularity}, leading to regularity constraints including those known as ``bubble equations'', and we examine the resulting structure of the metric near the bolts. Using these results, we describe some topological properties of these solutions and the three-form fluxes supporting the two bolts. 
	Section \ref{sec:examples-num} is devoted to exhibiting a set of examples obtained by solving the bubble equations for large collections of the integers parametrising our solutions. We discuss in detail solutions with an approximate AdS$_3$ region, as well as a family of solutions with parametrically small charge-to-mass ratio. 
	Appendix \ref{app:system} contains the explicit expressions for our supergravity ansatz; Appendix \ref{app:coord-change} discusses the adapted coordinates near the three special points; and finally, Appendix \ref{app:vec-fields} includes explicit expressions for the various vector fields arising in our solutions.

	\section{The supergravity construction}
	\label{sec:sugra-sol}
	
	In this section, we describe a partially-solvable system for constructing solutions to six-dimensional $\cN=(1, 0)$ supergravity coupled to a tensor multiplet. We then discuss briefly the structure of a three-centre gravitational instanton recently constructed in \cite{Chen:2011tc,Chen:2015vva}, which can be viewed as two Kerr-NUT bolts touching at a point. Finally, we construct a family of six dimensional supergravity solutions, using this instanton as a base for the partially-solvable system.
	
	\subsection{The partially-solvable system}
	
	We consider solutions to six-dimensional $\cN=(1, 0)$ supergravity coupled to a single tensor multiplet. The field content of this theory is the metric, a two-form potential, $C$, and a scalar field, $\phi$. These arise, for example, as a consistent truncation of Type IIB supergravity compactified on a Ricci-flat four dimensional manifold, and we will have in mind a D1-D5 bound state on T$^4$ or K3. In order to construct non-extremal solutions, we use the partially-solvable system of \cite{Bossard:2014ola}, in the formulation of \cite{Bena:2016dbw}, for six dimensional supergravity coupled to $n_T$ tensor multiplets. Here, we briefly introduce this system given in \cite{Bena:2016dbw} for the $n_T=1$ theory in which we work, referring the reader to that work for more details.
	
	The system of equations involves nine functions on a three-dimensional base space. Two of these functions, $V$, $\Vb$, determine the metric of the three-dimensional base space, as we describe momentarily. The seven other functions are organised in two triplets of functions, $K_I$, $L^I$, for $I=1,2,3$, and another function, $M$. These functions are organised into three layers, where each layer consists of linear second-order differential equations with sources that are determined by the data of the foregoing layers. These nine functions parametrise the full supergravity solution. 
	
	In more detail, the functions, $V$ and $\Vb$ satisfy the following Ernst equations and determine the three-dimensional base space metric, $\gamma_{ij}$, via
	\begin{gather}
	\Delta V = \frac{2\, \Vb}{1+ V\Vb}\,\nabla V\!\cdot\! \nabla V \,, \qquad 
	\Delta \Vb ~=~ \frac{2\, V}{1+V \Vb}\,\nabla \Vb\!\cdot\! \nabla \Vb\,, \cr
	R(\gamma)_{ij} = -\frac{ \partial_{(i\,}\!V \,  \partial_{j)}\!\Vb  }{( 1+V\Vb )^2} \,.
	\label{eq:R-base}
	\end{gather}
	Thus $V$ and $\Vb$ parametrise a four-dimensional gravitational instanton with one isometry. Note however that the corresponding four-dimensional Riemannian metric is auxiliary in the sense that it does not appear in the six-dimensional metric ansatz in our supergravity solutions. Furthermore, there is no physical coordinate associated to this isometry in the solutions we construct.

	For use below, we introduce the Hodge star, $\star$, and the Laplacian, $\Delta$, associated to the Riemannian metric $\gamma_{ij}$.
	The remaining equations take the following form:
	\begin{eqnarray}\label{eq:Lapl-eqns-gen}
	\Delta K_I &=& \frac{2\, V }{1+V \Vb}\,\nabla \Vb \!\cdot\! \nabla K_I\,, \cr
	\Delta L^I &=& \frac12\,\frac{V}{1+V \Vb}\,C^{IJK}\,\nabla  K_{J} \!\!\; \cdot\! \nabla K_{K} \,,  \\
	\Delta M  &=&  \nabla \!\cdot\!\left( \frac{V}{1+V\Vb}\, \left( L_I \nabla K^I -2\, M \nabla \Vb \right) \right) \,,
	\nonumber
	\end{eqnarray}
	where for the model at hand the structure constants, $C^{IJK}$, are given by symmetric permutations of $C^{123}=1$, and are zero otherwise. In the order written above, these equations are a solvable subsystem on the background specified by a solution $V,\?\Vb,\gamma_{ij}$ to \eqref{eq:R-base}. Note that there is a symmetry transformation leaving \eqref{eq:Lapl-eqns-gen} invariant: 
	\begin{align}
	K_I &\to K_I+ k_I \Vb\,, \cr 
	L^I &\to L^I + \frac12\,C^{IJK}\,k_J K_K + \frac14\,C^{IJK}\,k_J k_K \Vb\,, \cr
	M &\to M + \frac12\,k_I L^I + \frac18\,\frac{V}{1+V\Vb}\,C^{IJK}\,k_I K_J K_K \cr
	&\quad
	+ \frac14\,\left( 1- \frac12\,\frac{1}{1+V\Vb}\right)\, \left( C^{IJK}\,k_I k_J K_K + \frac13\,C^{IJK}\,k_I k_J k_K \Vb \right)\,,
	\label{eq:gaugespectral}
	\end{align}
	for some constants $k_I$. This symmetry is reminiscent of the gauge and spectral flow transformations present in solvable systems describing extremal solutions.

	Given a solution to the system \eqref{eq:Lapl-eqns-gen}, the six-dimensional metric, two-form and the scalar field that solve the supergravity equations of motion are as follows. We write the metric as
	\begin{equation}\label{eq:6D-metr}
	ds^2 = \frac{H_3}{ \sqrt{H_1\?H_2} } \?\left( d \yan + A^3 \right)^2 - \frac{W}{\sqrt{H_1\?H_2}\?H_3}\? \left( dt + k \right)^2 
	+ \sqrt{H_1\?H_2}\?\Scal{ \frac{1}{W} ( d\psi + w^0 )^2 + \gamma_{ij} dx^i dx^j} \,,
	\end{equation}
	in terms of a function, $W$, a vector of functions, $H_I$, and three one-forms, $A^3$, $k$ and $w^0$. Note the Kaluza--Klein structure, with $A^3$ the Kaluza--Klein gauge field, anticipating our focus on asymptotically-flat solutions in five dimensions. The forms $A^3$ and $k$ decompose as
	\begin{equation}\label{eq:6d-KK}
	A^3 ~=~ A^3_t\, (dt + \omega) + \ax^3\,(d\psi + w^0) +  w^3 \,, \qquad\quad k ~=~ \frac{\mu}{W}\,(d\psi + w^0)+\omega  \,,
	\end{equation}
	where $A^3_t$, $\ax^3$, $\mu$ and $w^0$, $\omega$ are respectively three scalars and two vector fields on the three-dimensional base.
	
	All functions appearing in the supergravity fields are given in terms of the functions  $(V,\Vb,K_I,L^I,M)$, as specified in \cite{Bena:2016dbw} and summarised here. The functions $W$, $\mu$, $H_I$ that appear in the metric are determined as follows:
	\begin{align}\label{eq:scal-facts}
	W = &\, \left(  (1 + \Vb)\, M - \frac12 \,\sum_{I=1}^3 K_I L^I +  \frac{1}{4}\,\frac{V}{1+V\Vb}\,K_1\? K_2\? K_3 \right)^2  \nonumber\\
	&+\frac{1-V}{1+V\Vb}\,\left( K_1\? K_2\? K_3\, M 
	+2\, (1 + \Vb)\,  L^1\? L^2\? L^3
	-\sum_{I=1}^3  K_{I+1} K_{I+2}\, L^{I+1} L^{I+2}  \right) \,,
	\nonumber\\
	H_I = &\, L^{I+1} L^{I+2} - K_I\, M + \frac12\,\frac{V}{1+V\Vb}\, (K_I )^2\,L^I  \,,
	\nonumber\\
	\mu = &\, (1 + \Vb)\,M^2 -\frac12\, M\, \sum_{I=1}^3 K_I L^I 
	- \left( 1 + 2\,\frac{V-1}{1+V\Vb} \right)\,L^1\? L^2\? L^3
	\CR
	&\, + \frac12\,\frac{V}{1+V\Vb}\,\left( -\frac1{2}\, K_1\? K_2\? K_3\, M +\sum_{I=1}^3  K_{I+1} K_{I+2}\, L^{I+1} L^{I+2} \right) \,,
	\end{align}
	where we use cyclic notation modulo $3$ for the index $I$.
	The corresponding expressions for the remaining quantities appearing in \eqref{eq:6D-metr}--\eqref{eq:6d-KK}, as well as all other quantities appearing in the expressions for the matter fields below, are given in Appendix \ref{app:system}.
	
	Turning to the matter fields, the scalar field, $\phi$, corresponding to the Type IIB dilaton, is given by
	\begin{equation}
	e^{2 \phi} = \frac{H_1}{H_2} \ .
	\end{equation}
	The two-form potential, $C$, is conveniently described in terms of two three-form field strengths $G_a=dC_a$, where the $C_a$ for $a=1,2$ denote the corresponding two-form potentials. These two field strengths satisfy the twisted self-duality equation
	\begin{equation} \label{eq:6D-eom}
	e^{\phi} \star_6 G_1 +e^{-\phi} G_2 = 0  \,. 
	\end{equation}
	This is equivalent to the equation of motion for the original two-form $C$. In this notation, the anti-self-dual combination of the three-form is part of the supergravity multiplet, while the self-dual combination is seen as part of the tensor multiplet, together with the dilaton. 
	We also define the invariant $SO(1,1)$ metric  
	\begin{equation}\label{eq:STU-eta}
	\eta_{ab}  = \left(\begin{array}{cc} 0\ &\ 1\\1\ &\ 0\end{array}\right) \,,
	\end{equation}
	and its inverse, $\eta^{ab}$, that is useful in writing explicit expressions below.
	
	In terms of three-dimensional quantities, the two-form potentials $C_a$ are decomposed in terms of scalars $A_t^a$, $\beta_a$ and $\ax^a$, where $\ax^a$ are identified as axions in the reduction to four dimensions, as well as three-dimensional one-forms $w^a$, $v_a$ and $b_a$. The explicit decomposition, assuming axisymmetry as we do, takes the form\footnote{In the general case without axisymmetry, there is an additional term, which can be found in \cite{Bena:2016dbw}.}
	\begin{align}
	C_a =&\, \eta_{ab}\? A_t^b\, (d\yan + w^3 ) \wedge(dt+\omega) + \eta_{ab}\? \ax^b\, (d\yan + w^3)  \wedge (d\psi + w^0)
	-\beta_a\,(dt+\omega)  \wedge (d\psi + w^0) \quad 
	\CR
	&\,- \eta_{ab}\? w^b \wedge (d\yan + w^3 )+ b_a \wedge(dt+\omega) + v_a \wedge (d\psi + w^0 )  \,.
	\label{eq:two-form-exp}
	\end{align}
	The relevant expressions for the fields $A_t^I$, $\ax^I$, $v_a$, $b_a$ and $\beta_a$ in terms of the functions solving the system \eqref{eq:R-base}--\eqref{eq:Lapl-eqns-gen} can be found in Appendix \ref{app:system}, and are used to compute these quantities throughout this paper.

	\subsection{Two-bolt gravitational instanton}
	\label{sec:two-bolt-inst}
	
	In order to construct supergravity solutions of the type described in the previous section, one starts with a gravitational instanton that solves the nonlinear part of the system in \eqref{eq:R-base}. An obvious example of such an instanton is the Kerr-NUT bolt, \ie the analytic continuation of the Kerr-NUT black hole solution to Euclidean signature, for which the horizon is replaced by a smooth bolt. Solutions based on this instanton and generalisations thereof were constructed in \cite{Bossard:2014yta, Bossard:2014ola, Bena:2015drs, Bena:2016dbw}.
	
	Obtaining more general solutions to the Euclidean Einstein equations with a single isometry, or even two commuting isometries, is a difficult task. For our purposes, a natural strategy is to take a known class of instantons solving the first layer of our system~\eqref{eq:R-base}, and use them as the basis for solving the remaining layers of equations to find new six-dimensional solutions. 
	
	In this paper, we consider the three-centre four-dimensional gravitational instanton of \cite{Chen:2015vva}, which can be thought of as describing two Kerr-NUT bolts, touching at a single point along their common axis of rotation. Note that this more general instanton does not admit an analytic continuation to Minkowski signature, since the gravitational attraction would of course lead the resulting pair of touching black holes to merge. Below, we describe the instanton metric and some of the conditions required for its regularity. Note that smoothness, including absence of conical singularities, imposes quantisation conditions on the parameters of the four-dimensional solution. However, we shall not describe these, since we are only interested in smoothness of the six-dimensional supergravity solution, and in general, the quantisation conditions are different. We discuss the smoothness analysis of the full six-dimensional supergravity solution in the next section. 
	
	\vspace{-1mm}
	\subsubsection{The metric}
	\vspace{-1mm}
	
	We now review the four-dimensional gravitational instanton of \cite{Chen:2015vva}.
	The solution is described in terms of a quartic polynomial, whose coefficients parametrise the solution, as
	\begin{align}\label{eq:inst-poly}
	P(u)=& \, a_0 + a_1\?u + a_2\?u^2 + a_3\?u^3 + a_4\?u^4 
	\CR
	=& \, a_4\?(u-t_1)\?(u-t_2)\?(u-t_3)\?(u-t_4)\,,
	\end{align}
	where either the $a_i$ or the roots $t_i$ can be used as parameters, and the relation between the two is given by
	\begin{gather}
	a_0 = a_4\?t_1\?t_2\?t_3\?t_4 \,, \quad a_1 = - a_4\?(t_1\?t_2\?t_3 + t_1\?t_2\?t_4 + t_1\?t_3\?t_4 + t_2\?t_3\?t_4) \,,\CR
	a_2 = a_4\?(t_1\?t_2 + t_1\?t_3 + t_1\?t_4 + t_2\?t_3 +t_2\?t_4 + t_3\?t_4) \, , \quad
	a_3 = - a_4\?(t_1 +t_2 + t_3 + t_4)\,.
	\label{eq:atr}
	\end{gather}
	
	As is natural for an axisymmetric gravitational instanton, the metric is written as a circle fibration, with coordinate $\tau$, over a three-dimensional base space metric that is independent of the angle, $\varphi$, around the axis of symmetry.
	The remaining two directions along the three-dimensional base are expressed in coordinates $x$, $y$, reminiscent of the adapted coordinate system used for black ring solutions.\footnote{Note that $y$ is completely distinct from the coordinate $\yan$ used in the six-dimensional metric \eq{eq:6D-metr}.} The polynomial \eqref{eq:inst-poly} appears symmetrically for the $x$ and $y$ coordinates, as
	\begin{align}\label{eq:XY-def}
	X= \, P(x) \,, \qquad Y= \, - P(y) \,,
	\end{align}
	in terms of which the metric is given by
	\begin{equation}\label{eq:inst-metric}
	ds_4^2 = \frac{F}{(x-y)\?H} \left(d\tau +\frac{G}{F}\? d\varphi \right)^2
	+ \frac{H}{(x-y)^3\?F}\? \left(\kappa^2\? F\?
	\left(\frac{dx^2}{X}+\frac{dy^2}{Y}\right)+X\? Y\?d\varphi^2 \right) \,,
	\end{equation}
	where
	\begin{align}
	F(x,y) =&\, x^2\? Y + y^2\? X \,, \cr
	H(x,y)=&\, (\nu\? x+y) \bigl[ (\nu\? x-y) (a_1- a_3\? x\? y)-2\? (1-\nu ) \left( a_0 - a_4\? x^2\? y^2\right)\bigr] , \\
	G(x,y)=&\, X \? \left[ \nu^2\? a_0 +2\? \nu\? a_3\? y^3+( 2\? \nu - 1 )\? a_4\? y^4 \right] 
	-Y \? \left[ ( 1 -2\? \nu\ )\? a_0 - 2\?\nu\? a_1\? x - \nu^2\? a_4\?  x^4\right]\,,
	\nonumber
	\end{align}
	and where $\kappa$ and $\nu$ are  two additional constant parameters. 
	In this form, where the fibration is along the coordinate $\tau$, we can read off the relevant Ernst potentials describing the solution through their definition
	\begin{equation}
	\cE_+ + \cE_- ~=~  \frac{F}{(x-y)\?H}  \,, \qquad \quad
	d(\cE_+ - \cE_- ) ~=~ \frac{(x-y)^2\?H^2}{F^2} \star d\left(  \frac{G}{F}\? d\varphi \right) \,,
	\end{equation}
	which leads to the explicit expressions
	\begin{equation} \label{ErnstOriginal}
	\cE_+ ~=~ \frac{(x-y)\? (\nu\? x+y) (a_1 - a_3\? x\? y)}{2\? (\nu -1)\?H} \,, \qquad
	\cE_-  ~=~ -\frac{x+y}{2\? (\nu -1)\? (\nu\? x+y)} \,.
	\end{equation}
	One can verify that $\cE_{\pm}$ indeed satisfy the Ernst equations,
	\begin{equation}\label{eq:Ernst}
	\left( \cE_+ + \cE_- \right)\?\Delta \cE_\pm = 2\, \nabla \cE_\pm\!\cdot\! \nabla \cE_\pm \,.
	\end{equation}
	Furthermore, it is clear from inspection of \eqref{ErnstOriginal} that there is no complex choice of the parameters and coordinates that would result in the relation $\cE_- = \overline{\cE_+}$, as would be required for a Minkowski-signature solution, confirming the fact that this solution does not admit an analytic continuation to a real pseudo-Riemannian metric. 
	
	In view of the presence of the additional isometry along $\varphi$, one can bring the metric to the canonical Weyl form
	\begin{equation}\label{eq:Weyl-form}
	ds_4^2 ~=~ \left( \cE_+ + \cE_- \right)\,\left(d\tau +\frac{G}{F}\? d\varphi \right)^2
	+\left( \cE_+ + \cE_- \right)^{-1}\left[ e^{2\sigma} \left( dz^2+d\rho^2 \right) +\rho^2 d\varphi^2 \right] \,,
	\end{equation}
	where the Weyl coordinates $\rho$, $z$ are related to the $x$, $y$ coordinates as
	\begin{equation}\label{eq:Weyl-coord}
	\rho^2 ~=~ \frac{X\? Y}{(x-y)^4} \,, \qquad 
	z~=~ \frac{2\?(a_0 + a_2\? x\? y + a_4\? x^2\? y^2) + (x + y)\?(a_1 + a_3\? x\? y)}{2\,(x-y)^2} \,,
	\end{equation}
	while the three-dimensional base space metric $\gamma_{ij} dx^i dx^j = e^{2\sigma} \left( dz^2+d\rho^2 \right) +\rho^2 d\varphi^2$ is described by the function
	\begin{equation}
	e^{2\sigma} ~=~  \frac{\kappa^2\?F}{(x-y)^4} \,.
	\end{equation}
	
	Since a zero of the polynomial in \eqref{eq:inst-poly} results in a divergent term in the metric \eqref{eq:inst-metric} through either $X$ or $Y$ in \eqref{eq:XY-def}, each of the coordinates $x$, $y$ may only take values in a range between two distinct, neighbouring roots of this quartic polynomial. Moreover, since we need both $X>0$ and $Y>0$ for a regular metric and since asymptotic infinity is reached by taking both $x$ and $y$ to be equal to one of the roots, it follows from \eqref{eq:XY-def} that $x$, $y$ must take values in two {\it adjacent} ranges, which are then specified by three of the roots of the polynomial~\cite{Chen:2015vva}. We choose by convention to order the roots $t_1$, $t_2$, $t_3$, so that we have
	\begin{equation}\label{eq:root-order}
	t_1\leq x \leq t_2 \leq y \leq t_3 \,, \quad \text{or}  \quad  t_3\leq y \leq t_2 \leq x \leq t_1\,,
	\end{equation}
	and so that $t_4$ is outside this range, \ie either greater than $t_3$ or smaller than $t_1$ in the first case, or smaller than $t_3$ or greater than $t_1$ in the second.\footnote{In general one could also consider a range that includes infinity, but we shall not do this in this paper.} Through \eqref{eq:Weyl-coord} the above domain \eq{eq:root-order} is homeomorphic to the half-plane of the Weyl coordinates $(\rho,z)$, \ie $\rho\ge 0$ and $z\in \mathds{R}$, as we explain in Appendix \ref{app:coord-change}.

	The solution contains three special points in its interior; these are reached when both $x$ and $y$ reach either boundary of their respective ranges. The locations $z_i$ of these points along the axis of symmetry are given in terms of the roots of the polynomial as follows:
	\begin{align}\label{eq:points}
	\text{Asympt. infinity: }&\,\quad     x = t_2\,, \ \  y = t_2 \,, \,\quad  
	\CR
	\text{Centre 1: }&\,\quad               x = t_1\,, \ \  y = t_2 \,, \,\quad    z_1 =-\tfrac12\,a_4\, (t_1 t_2 + t_3 t_4) \,,
	\CR
	\text{Centre 2: }&\,\quad               x = t_1\,, \ \  y = t_3 \,, \,\quad    z_2 =-\tfrac12\,a_4\, (t_1 t_3 + t_2 t_4) \,,
	\CR
	\text{Centre 3: }&\,\quad               x = t_2\,, \ \  y = t_3 \,, \,\quad    z_3 =-\tfrac12\,a_4\, (t_1 t_4 + t_2 t_3) \,.
	\end{align}
	For later reference, we define local spherical coordinates close to each of these four loci, namely $(r\,, \cos\theta)$ for asymptotic infinity and $(r_i\,, \cos\theta_i)$ for each of the centres at $z_1$, $z_2$, $z_3$, as follows:
	\begin{alignat}{2}\label{eq:Weyl-to-r}
	r =&\, \sqrt{\rho^2 + z^2}\,,\qquad  && \cos\theta = \frac{z}{\sqrt{\rho^2 + z^2}}\,,
	\CR
	r_i =&\, \sqrt{\rho^2 + (z-z_i)^2}\,,\qquad &&\cos\theta_i = \frac{z-z_i}{\sqrt{\rho^2 + (z-z_i)^2}}\,,
	\end{alignat}
	where $\rho$, $z$ are given by \eqref{eq:Weyl-coord}, with the choice of branch for the roots such that the special points satisfy \eqref{eq:points}.
	
	Parametrisation invariance induces the unphysical rescaling symmetry of the system
	\be \label{Reparaxy}  (x,y)\rightarrow (\lambda x,\lambda y) \ , \qquad a_i \rightarrow \lambda^{2-i} a_i \ , \qquad t_i \rightarrow \lambda t_i \ , \ee
	that leaves invariant the components of the metric in Weyl coordinates. One could therefore fix this redundancy by choosing for example $a_4=1$. However it will be more convenient for us to keep all parameters free at this stage, and to fix this redundancy at a later point.

	\vspace{-1mm}
	\subsubsection{Supergravity embedding}
	\label{sec:base-infinity}
	\vspace{-1mm}
	
	In order to use this instanton in the partially-solvable system of the previous section, one needs to identify appropriate functions $V$ and $\Vb$ solving \eqref{eq:R-base}. These are equivalent to the Ernst equations \eqref{eq:Ernst}, so that $\Vb$ and $V^{-1}$ can be identified with the $\cE_\pm$. However the Ernst equations are invariant under the exchange of $\cE_+$ and $\cE_-$, and also under $SL(2)$ real fractional linear transformations, so there is some freedom in this identification. We fix this freedom by choosing $V$ to be the simpler of the two functions, and thus related to $\cE_-$ by a fractional linear transformation. Then $\Vb$ is related to $\cE_+$ in a similar fashion, explicitly:
	\begin{eqnarray}\label{eq:VVb-gen}
	\Vb(\cE_+) &=&\frac{ \alpha \cE_+ + \beta}{\gamma \cE_+ + \delta} \,, \quad\quad
	V(\cE_-) ~=~  -\frac{\gamma \cE_- -\delta}{\alpha \cE_- -\beta } \,.
	\end{eqnarray}
	The invariance under $SL(2)$ comes from the fact that the Ernst equations  \eqref{eq:Ernst} describe a non-linear sigma model over $SL(2)/SO(1,1)$. The choice of parameters in  \eqref{eq:VVb-gen} determines the asymptotic fall-off, and we shall use it to obtain the appropriate asymptotic behaviour of the functions $V$ and $\Vb$ needed to obtain asymptotically $\mathds{R}^{1,4} \times S^1$ supergravity solutions.
	
	Upon expanding the metric \eqref{eq:inst-metric} around asymptotic infinity using \eqref{eq:asym-expansions}, we find that the three-dimensional base becomes the flat metric on $\mathbb{R}^3$ for the choice
	\begin{equation}\label{eq:kappa-val}
	\kappa = \frac{a_4\? (t_2 - t_1)\? (t_2 - t_3)\? (t_2 - t_4)}{2\, t_2 }\,.
	\end{equation}
	Similarly, expanding the functions \eqref{eq:VVb-gen} and imposing that $V$ asymptotes to $1$ and that $\Vb$ vanishes asymptotically, as for the similar three-centre instanton in \cite{Bena:2016dbw}, we restrict the parameters of the $SL(2)$ transformation to be given by
	\begin{equation}
	\alpha = 1\,, \quad\beta = 0\,, \quad 
	\gamma = -1 -\frac{\nu +1}{\nu -1}\,\hat\delta\,, \quad \delta = \frac{\hat\delta }{(\nu -1)^2}\,.
	\end{equation}
	The rescaled parameter, $\hat\delta$, remains free and replaces the original parameter, $\nu$, which does not appear in the expressions for $V$, $\Vb$. Having done this rescaling, we now drop the hat from $\hat\delta$ for notational convenience.
	Explicitly, $V$ and $\Vb$ then become:
	\begin{align}\label{eq:VVb-fin}
	V =&\, 1-\delta\, \frac{x-y}{x+y} \,, \cr
	\Vb =&\,  \frac{(x-y) (a_1- a_3\? x\? y)}{4\? \delta\?  \left({a_0}-{a_4}\? x^2\? y^2\right) + (a_1- a_3\? x\? y)\? \left((\delta -1) x+(\delta +1) y\right)} \,. \qquad
	\end{align}
	
	With this choice of $V$, $\Vb$, the relevant scale factor of the metric, $\left( 1 + V\, \Vb \right)^{-1}$, contains simple poles at the three centres, as one may verify by expanding the coordinates $(x,y)$ to the first nontrivial order in the spherical coordinates \eqref{eq:Weyl-to-r} around each special point; these expansions are given in \eqref{eq:asym-expansions}. The corresponding coefficients are proportional to the three combinations of parameters
	\begin{equation}\label{eq:ni}
	p_1 = \, t_1 t_2 - t_3 t_4 \,, \qquad p_2 = \, t_1 t_3 - t_2 t_4 \,, \qquad p_3 =\,t_1 t_4 - t_2 t_3\,, 
	\end{equation}
	which, together with $\kappa$ given in \eqref{eq:kappa-val}, will be useful shorthands in the various expressions of the supergravity solutions to be constructed in the next section. 
	
	As was shown in \cite{Chen:2015vva}, there exists a limit in which the metric \eqref{eq:inst-metric} reduces to a three-centre Gibbons--Hawking instanton with charges proportional to \eqref{eq:ni}.
	
	\vspace{-1mm}
	\subsubsection{The three-dimensional base metric near the nuts and bolts}
	\label{sec:base-bolts}
	\vspace{-1mm}
	
	As described above, the three-dimensional base of the above four-dimensional gravitational instanton will be the three-dimensional base of the six-dimensional solutions we construct. 
	We now examine the form of this base metric near the special loci, using the local coordinates defined above. From \eqref{eq:inst-metric} and \eqref{eq:Weyl-form}, the three-dimensional base metric is given by 
	\begin{equation}\label{eq:3d-base}
	ds_3^2 
	~=~  \frac{\kappa^2\? F}{(x-y)^4} \left(\frac{dx^2}{X}+\frac{dy^2}{Y}\right)+\frac{X\? Y}{(x-y)^4}\?d\varphi^2
	~=~  e^{2\sigma} \left( dz^2+d\rho^2 \right) +\rho^2 d\varphi^2 \,.
	\end{equation}
	In the adapted Weyl coordinates, all the special loci are located along the axis of symmetry, at $\rho=0$. 
	
	Starting from the three special points $z=z_i$ in \eqref{eq:points}, we change to the adapted spherical coordinates, $r_i$, in \eqref{eq:Weyl-to-r}, and use the local expansions given in \eqref{eq:asym-expansions} to find the following form of the metric near each point:
	\begin{equation}\label{eq:base-centres}
	ds_3^2\?\Bigr|_{z_i}\!\!=~ e^{2\sigma_i} \left( d r_i^2 + r_i^2 d\theta_i^2 \right)+ r_i^2 \sin^2{ \theta_i} d \varphi^2 + {\cal O}(r_i^3)\,,
	\end{equation}
	where the functions $e^{2\sigma_i}$ are given by
	\begin{align} \label{sigmaati}
	e^{2\sigma_1} =&\, \kappa^2\,\left( \frac12\,(1 + \cos{ \theta_1}) +\frac12\,b_\pA^2\, (1 - \cos{ \theta_1})  \right) \,,
	\CR
	e^{2\sigma_2} =&\, \kappa^2\,\left( \frac12\,b_\pA^2\,(1 + \cos{ \theta_2}) +\frac12\,b_\pB^2\, (1 - \cos{ \theta_2})  \right) \,,
	\\
	e^{2\sigma_3} =&\, \kappa^2\,\left( \frac12\,b_\pB^2\,(1 + \cos{ \theta_3}) +\frac12\, (1 - \cos{ \theta_3})  \right) \,,
	\nonumber
	\end{align}
	and where we use the shorthand constants
	\begin{equation}\label{eq:bAB-def}
	b_\pA \,\equiv\, \frac{t_1\? (t_2 - t_3)\? (t_2 - t_4)}{t_2\? (t_1 - t_3)\? (t_1 - t_4)}\,, 
	\qquad 
	b_\pB \,\equiv\, \frac{t_3\? (t_1 - t_2)\? (t_2 - t_4)}{t_2\? (t_1 - t_3)\? (t_3 - t_4)}\,, 
	\end{equation}
	whose significance will become clear shortly.
	
	The two regions between the three special points were interpreted in \cite{Chen:2015vva} as two Kerr-NUT bolts with a common axis of rotation, touching at the nut at the middle point. With our choice of ordering for the roots in \eqref{eq:root-order}, the position in the middle where the two Kerr-NUT bolts touch is always $z_2$. Without loss of generality, we assume the ordering $z_3<z_2<z_1$, so that the half-lengths of the segments supporting the two four-dimensional bolts are 
	\begin{equation}\label{eq:cAB-def}
	c_\pA \,\equiv\, \frac12\,(z_1 - z_2) \,=\, - \frac{a_4}4\,(t_1 - t_4)\, (t_2 - t_3) \,, 
	\quad 
	c_\pB \,\equiv\, \frac12\,(z_2 - z_3) \,=\, -\frac{a_4}4\,(t_1 - t_2)\, (t_3 - t_4) \,. 
	\end{equation}
	
	Since we only use the three-dimensional base of the gravitational instanton, the interpretation of the two four-dimensional bolts is a priori lost, and one must perform the analogous analysis on the final six-dimensional supergravity solution. Nevertheless, it is useful to consider the metric near these two regions, as it will be useful preparation for the supergravity analysis in the next section. 
	We therefore define adapted radial and angular coordinates around each segment, as
	\begin{align}\label{eq:rAB-def}
	r_\pA = &\, \frac12\,(r_1 + r_2) \,, \qquad \cos\theta_\pA = \frac1{2\,c_\pA}\,(r_1 - r_2) \,,
	\CR
	r_\pB = &\, \frac12\,(r_2 + r_3) \,, \qquad \cos\theta_\pB = \frac1{2\,c_\pB}\,(r_2 - r_3) \,.
	\end{align}
	In terms of these, the base metric can be expanded around $r_\pA=c_\pA$ and $r_\pB=c_\pB$, leading to the expressions
	\begin{align}\label{eq:base-bolt}
	ds_3^2\?\Bigr|_{r_\pA}\!\!=~\! \scal{ r_\pA^2 - c_\pA^2 + b_\pA^2 c_\pA^2 \sin^2\theta_\pA} \left( \frac{dr_\pA^2}{r_\pA^2 - c_\pA^2  } + d\theta_\pA^2 \right) 
	+  \scal{ r_\pA^2 - c_\pA^2 } \sin^2\theta_\pA\, d\varphi^2 + {\cal O}(r_\pA^3)\,,
	\CR
	ds_3^2\?\Bigr|_{r_\pB}\!\!=~\! \scal{ r_\pB^2 - c_\pA^2 + b_\pB^2 c_\pB^2 \sin^2\theta_\pB} \left( \frac{dr_\pB^2}{r_\pB^2 - c_\pB^2  } + d\theta_\pB^2 \right) 
	+  \scal{ r_\pB^2 - c_\pB^2 } \sin^2\theta_\pB\, d\varphi^2 + {\cal O}(r_\pB^3)\,,
	\end{align}
	where we used the value of $\kappa$ given in \eqref{eq:kappa-val}. This local form of the base metric
	near each segment is exactly the same as for a Kerr-NUT instanton, with non-extremality parameters $c_\pA \sqrt{ b_\pA^{\; 2} -1} $, $c_\pB\sqrt{ b_\pB^{\; 2} -1}$ and `rotation' parameters $b_\pA c_\pA$, $b_\pB c_\pB$, respectively, thus justifying the definitions of $b_\pA$, $b_\pB$ in \eqref{eq:bAB-def}. The solution is extremal when either $b_\pA =\pm 1$ or $c_\pA = 0$ and  either $b_\pB =\pm 1$ or $c_\pB = 0$.

	\subsection{Two-bolt six-dimensional supergravity solution}
	\label{sec:solution}
	
	We now proceed to construct an explicit family of supergravity solutions based on the two-bolt four-dimensional gravitational instanton. This involves solving the remaining equations of our system \eqref{eq:Lapl-eqns-gen} on the three-dimensional base space \eqref{eq:3d-base} described in the previous section. Obtaining the general solution to this system is beyond the scope of this paper, as it would involve introducing functions with poles away from the bolts, similar to the construction of \cite{Bena:2016dbw} where additional Gibbons--Hawking centres were added to a single Kerr-NUT instanton. In this paper we restrict attention to solutions where the functions $K^I$, $L_I$, $M$ do not contain additional poles. By doing so, one can assume that the solution can be expressed in terms of rational functions of $x$ and $y$.
	
	The simplest solution of the remaining equations of our system \eqref{eq:Lapl-eqns-gen}, based on any given instanton, arises by acting with the symmetry operation in \eqref{eq:gaugespectral} on the trivial solution in which all the remaining functions $K^I$, $L_I$, $M$ vanish. This solution depends only on the two functions $V$ and $\Vb$ that define the gravitational instanton. However, it turns out that this solution is not general enough to include interesting solutions, so we must obtain a non-trivial solution to the subsystem \eqref{eq:Lapl-eqns-gen} in this background. As mentioned above, we assume that the functions are rational in $(x,y)$ coordinates. For extremal solutions, harmonic functions with poles of high degree tend to produce singular solutions, so we seek a non-trivial solution with the least possible singular behaviour at the special points. We find the following solution for the $K_I$  (which is regular at the special points):
	\begin{equation}
	K_I= m_I \,K_0 \,, \qquad K_0 =  -\frac{ x-y }{4\? \delta\?  \left({a_0}-{a_4}\? x^2\? y^2\right) + (a_1- a_3\? x\? y)\? ((\delta -1) x+(\delta +1) y)} \,,
	\end{equation}
	for three constants $m_I$. Given this solution, one can solve the remaining equations with the resulting source. 
	We now act with \eqref{eq:gaugespectral} on this solution and describe the resulting family of solutions.
	We thus take the solution for the $K_I$ to be given by
	\begin{equation}\label{eq:KI-def}
	K_I \equiv h_I + \tilde{K}_I = h_I + m_I\?K_0 + k_I\? \Vb\,,
	\end{equation}
	where the $h_I$ and $m_I$ are two triplets of constants. Note that we also defined the functions $\tilde{K}_I$, which do not include the asymptotic constants, $h_I$, and which will be useful in writing the remaining functions in the solution.
	
	The solution for the remaining functions can then be computed straightforwardly, by solving these equations for $k_I=0$ in the source terms and then using \eq{eq:gaugespectral} to re-introduce the dependence on the $k_I$. The resulting solution for the $L^I$ takes the form
	\begin{equation}\label{eq:LI-def}
	L^I = l^I +\frac{1}{4\,\Vb}\?C^{IJK} \tilde{K}_{J} \tilde{K}_{K} 
	-\frac{a_1^2\? a_3^2}{16\?(a_0\? a_3^2 - a_1^2\? a_4)\? \delta}\?C^{IJK} m_{J} m_{K}\?\frac{x-y}{a_1 - a_3\? x\? y} \,,
	\end{equation}
	where $l^I$ are a triplet of constants, and where the term quadratic in the $\tilde{K}_I$ includes all terms that depend on $k_I$ and can be seen to reproduce the dependence in \eq{eq:gaugespectral}.
	Similar comments apply to the $k_I$-dependent terms in $M$, for which we find the solution:
	\begin{align}\label{eq:M-def}
	M=&\, \frac{1}{2\,\Vb}\?\tilde{K}_{I} L^I 
	- \frac{1}{4\?\Vb^2}\,\frac{2+ V\,\Vb}{1+ V\,\Vb}\? \tilde{K}_{1}\tilde{K}_{2}\tilde{K}_{3}
	+ \frac{1}{1+V\?\Vb}\,\left( l_0 + q_0 (1 - V) - \frac{1}{2\?\Vb}\?l^I \tilde{K}_{I}\right)
	\CR
	-&\, \frac{a_1^3\? a_3^3}{16\?(a_0\? a_3^2 - a_1^2\? a_4)^2\? \delta^2}\?
	\frac{(1-V)}{1+V\?\Vb}\? \frac{m_{1}\? m_{2}\? m_{3}}{a_1 - a_3\? x\? y}\?
	\left( a_1 + \frac{2\? (x + y)\? (a_0\? a_3 - a_1\? a_4\? x\? y)}{a_1 - a_3\? x\? y}
	\right)\,,
	\end{align}
	where $l_0$ and $q_0$ are constants parametrising two homogeneous solutions of the last equation in \eqref{eq:Lapl-eqns-gen} that are generic for any gravitational instanton, as they are the same functions of $V$ and $\Vb$ for any solution to the Ernst equations \eqref{eq:R-base}.
	
	\section{Regularity of the solutions}
	\label{sec:regularity}
	
	In this section, we analyse some general properties and specify the constraints arising from regularity for the family of solutions found in the previous section. The analysis proceeds in three steps: we study asymptotic infinity in Section \ref{sec:infinity}, then we examine the structure and regularity of the solution near the three centres in Section \ref{sec:nuts}, and around the two bolts in Section \ref{sec:bolts}.
	
	\subsection{Asymptotic structure}
	\label{sec:infinity}
	
	We start by analyzing the asymptotic region and imposing the appropriate fall-off behaviour for the ansatz quantities, such that the resulting solutions are asymptotically $\mathds{R}^{1,4} \times $S$^1$.
	
	\vspace{-1mm}
	\subsubsection*{Useful redefinitions}
	\vspace{-1mm}
	
	In order to analyse the structure of the solution near asymptotic infinity, we first make a set of gauge transformations and coordinate transformations on the solution that is obtained by directly substituting \eqref{eq:KI-def}--\eqref{eq:M-def} in the relevant expressions, in order to obtain standard asymptotic values for the various fields. These operations do not impose any constraints on the parameters of the general solution, but represent a choice of gauge that we exploit in setting various asymptotic constants to zero.
	
	We first shift away the asymptotic constants from the off-diagonal components of the metric and the two-forms $C_a$. Concretely, one can shift to zero the asymptotic values of the scalars $\ax^a$, $\beta_a$ and $A_t^a$ in \eqref{eq:two-form-exp} by performing a gauge transformation on the two-forms, leading to the following redefinition of the vector fields
	\begin{align}\label{eq:redef1}
	w^a{}' = &\; w^a + A_t^a \big|_{\scriptscriptstyle\infty}\omega + \ax^a \big|_{\scriptscriptstyle\infty} w^0\ , 
	\CR
	v_a' = &\; v_a - \beta_a \big|_{\scriptscriptstyle\infty} \omega 
	+ \eta_{ab}\,\ax^b \big|_{\scriptscriptstyle\infty} w^3 \ , 
	\CR
	b_a' = &\; b_a + \eta_{ab}\,A_t^b \big|_{\scriptscriptstyle\infty} w^3
	+ \beta_a \big|_{\scriptscriptstyle\infty} w^0\ .
	\end{align}
	Here, primes denote redefined quantities, while asymptotic values are denoted by $\big|_{\scriptscriptstyle\infty}$,  and we use the first Pauli matrix $\eta_{ab}$ \eqref{eq:STU-eta}. Having made these redefinitions, we immediately drop the primes on the new quantities, and we will do likewise for the following two steps.
	
	In the same way we set to zero the asymptotic values of $A_t^3$ and $\ax^3$ that appear in the Kaluza--Klein gauge field $A^3$ given in \eqref{eq:6d-KK}, by performing a diffeomorphism that mixes the coordinate $\yan$ with $t$ and $\psi$ at infinity, leading to the following redefinitions
	\begin{align}\label{eq:redef2}
	v_a' = &\; v_a + \ax^3 \big|_{\scriptscriptstyle\infty}\eta_{ab}\, w^b \ , 
	\CR
	b_a' = &\; b_a + A_t^3 \big|_{\scriptscriptstyle\infty} \eta_{ab}\,w^b  \ , 
	\CR
	\beta_a'  = &\; \beta_a - \ax^3\big|_{\scriptscriptstyle\infty} \eta_{ab}\, A_t^b 
	+ A_t^3 \big|_{\scriptscriptstyle\infty} \eta_{ab}\,\ax^b\ . 
	\end{align}
	Moreover, one can shift away the asymptotic constant values of $\omega$, $w^3$ and the $w^a$ by making an appropriate diffeomorphism mixing the coordinates $t$, $\yan$ with $\varphi$, as well as by a further gauge transformation on the two-forms; we therefore assume that these vector fields vanish at infinity for the remainder of the paper. 
	
	Finally, we consider the freedom of choosing the asymptotic time coordinate as a linear combination of the coordinate $t$ with one of the compact directions, $t = t' + \gamma\, \psi$, leading to the  field redefinitions
	\begin{eqnarray}\label{eq:t-psi-mix}
	\omega' = \omega - \gamma\, w^0 \,, \qquad 
	\mu' = \mu + \gamma\, W \,,\qquad 
	\alpha^I{}' = \alpha^I + \gamma\, A_t^I\,,\qquad 
	v_I' = v_I + \gamma\, b_I \,.
	\end{eqnarray}
	This freedom, parametrised by the constant $\gamma$, will be fixed by imposing the asymptotic conditions below.
	We once again immediately drop the primes on all the above redefined quantities.
	
	The explicit expressions for the asymptotic constants appearing in the above redefinitions \eqref{eq:redef1}, \eqref{eq:redef2} are straightforward to obtain using the solution given in Section \ref{sec:solution}. However, these are not illuminating and play no role in the following, so we refrain from displaying them and henceforth work with the quantities obtained after \eqref{eq:redef1}--\eqref{eq:t-psi-mix} have been applied.
	
	\vspace{-1.5mm}
	\subsubsection*{Constraints on the asymptotic fall-off}
	\vspace{-1.5mm}
	
	We now turn to the constraints imposed by demanding that the asymptotic structure of the solution is that of a five-dimensional black hole. Starting from the metric, in order to obtain the desired $\mathds{R}^{4,1}\times $S$^1$ asymptotics, we impose that the functions that appear in the metric fall off as
	\begin{equation}\label{eq:as-beh1}
	W = \frac{1}{r^2}+ 2\?\xi_{\infty}\?\frac{\cos\theta}{r^3}+ {\cal O}\scal{r^{-4}}\,, \qquad 
	H_I = \frac{1}{r} + \frac{\frac{E_I}{4} + \xi_{\infty}\?\cos\theta}{r^2} + {\cal O}\scal{r^{-3}}\,, 
	\end{equation}
	where the $E_I$ are positive constants that parametrise the mass and the fall-off of the scalar fields at infinity, and where  $\xi_{\infty}$ is a real parameter. For the sake of simplicity, we follow the same approach as in \cite{Bena:2015drs, Bena:2016dbw} and fix the asymptotic values of $g_{\yan\yan}$ and the dilaton by fixing the leading fall-off of the $H_I$ functions to be $\frac{1}{r}$. There is no loss of generality arising from this choice, since we keep the radius of the $\yan$ circle general through its coordinate length $\yan \ident \yan+2\pi R_\yan\?$; similarly, the asymptotic value of the dilaton $e^{2\phi}|_\infty$ can straightforwardly be scaled to an arbitrary value using the global $GL(1)$ symmetry of the theory in six dimensions. 
	
	Next, we turn to the off-diagonal components of the metric \eqref{eq:6D-metr} that involve the time coordinate, and impose that they fall off at large $r$ as
	\begin{equation}\label{eq:as-beh2}
	\mu= \frac{-J_\psi + J_\varphi \, \cos{\theta}}{8 \, r^3} + {\cal O}(r^{-4})\,,
	\qquad
	\omega = -\frac{J_\varphi\, \sin^2{\theta}}{8 \, r} d\varphi+ {\cal O}(r^{ -2})\,,
	\end{equation}
	where $J_\psi$, $J_\varphi $ are the angular momenta along the corresponding angular directions.

	Finally, the asymptotically-flat five-dimensional solution that is obtained by reduction on the asymptotic circle has electric charges, $Q_I$, which are defined in terms of the fall-off of the time components of the five-dimensional gauge potentials, $A^I_t$, as
	\begin{equation}\label{eq:Q-def}
	A^I_t =  \frac{Q_I}{4 r}+ {\cal O}(r^{ -2})\,.
	\end{equation}
	Since we are interested in black hole microstate geometries, we wish to constrain the behaviour of the matter fields to that of an asymptotically-flat black hole solution. This requirement imposes the following relation between the electric charges and the quantities $E_I$ defined in \eqref{eq:as-beh1}:
	\begin{equation}\label{eq:as-beh3}
	E_1^2 - Q_1^2 \;=\; E_2^2 - Q_2^2 \;=\; E_3^2 - Q_3^2\; .
	\end{equation}
	This constraint can be understood as an attractor equation for the two scalar fields in five dimensions, such that their asymptotic momenta are determined by the electric charges and the mass of the black hole.\footnote{Note that the two five-dimensional scalar fields are parametrised by ratios of the $H_I$ functions, and behave as $\frac{H_I}{H_J} = 1+ \frac{E_I-E_J}{4\?r} + \mathcal{O}(r^{-2})$ at infinity. This condition is enforced in \cite{Cvetic:1996kv,Cvetic:2011hp}, because $E_I$ and $Q_I$ are proportional to the hyperbolic cosines and sines of the corresponding boost/duality parameters $\delta_I$.}

	In order to impose the conditions \eqref{eq:as-beh1}--\eqref{eq:as-beh3}, one must expand the solution obtained in Section \ref{sec:solution} near infinity and constrain some of the free parameters. Imposing the asymptotic behaviour \eqref{eq:as-beh1} results in fixing the values of the parameters $h_I$, $l^I$, $l_0$ and $q_0$ introduced in \eq{eq:KI-def}--\eq{eq:M-def}. We have
	\begin{gather}\label{eq:AsymPar0}
	l_0 = l^I  = \frac{1}{2}\,,
	\qquad
	h_I = 1 \, ,
	\end{gather}
	and we shall give the value of $q_0$ shortly, once we have introduced some more notation. Once this is done, one finds that $\mu =-\gamma/r^2 + {\cal O}\scal{r^{-3}}$ for a constant $\gamma$ that will be given below; we thus apply the transformation \eqref{eq:t-psi-mix} in order to remove the $ {\cal O}\scal{r^{-2}}$ term, obtaining the desired $ {\cal O}\scal{r^{-3}}$ fall-off in \eqref{eq:as-beh2}. The final conditions to consider are \eqref{eq:as-beh3}, which read
	\begin{align}
	E_1^2 - Q_1^2 \;=\; E_2^2 - Q_2^2 \quad~~ \Rightarrow \quad~~ &\, \frac{a_1^2\,a_3^2}{2\,a_4^3\,\delta}\, \frac{\kappa^2}{p_1\?p_2\?p_3}\,m_3  \left( m_1\? (k_2-1) - m_2\? (k_1-1) \right) \;=\; 0\,,
	\CR
	E_3^2 - Q_3^2 \;=\; E_2^2 - Q_2^2 \quad~~ \Rightarrow \quad~~ &\, \frac{a_1^2\,a_3^2}{2\,a_4^3\,\delta}\, \frac{\kappa^2}{p_1\?p_2\?p_3}\,m_1 \left( m_3\? (k_2-1) - m_2\? (k_3-1) \right) \;=\; 0\,,
	\end{align}
	where the constants $\kappa$ and $p_1$, $p_2$, $p_3\,$ are defined in \eqref{eq:kappa-val} and \eqref{eq:ni} respectively. Since setting $\kappa$ or either of $a_1$ or $a_3$ to zero would lead to a degenerate metric, we consider appropriate restrictions of the $m_I$. This leads to two branches of solutions to \eqref{eq:as-beh3}:
	\begin{align}\label{eq:rank13}
	\text{Rank 1:}\quad~~ &\, m_1 = m_2 =0\,, \quad m_3\neq 0 \,,
	\CR
	\text{Rank 3:}\quad~~ &\, m_I = m \?(k_I-1) \,,
	\end{align}
	where $m$ is an arbitrary constant, and we name each branch by the number of nonzero components. Note that the Rank 1 branch may be along any of the three directions, but we choose $m_3\neq 0$ in order to keep manifest the symmetry between the two two-form potentials \eqref{eq:two-form-exp}, since it is the vector field $A^3$ that appears in the six-dimensional metric \eqref{eq:6D-metr}. The other choices with only $m_1$ or $m_2$ non-zero define a-priori independent solutions that we have not studied. Henceforth, we will only consider the Rank 1 solution with $m_3\ne 0$ in \eqref{eq:rank13}, and we leave the analysis of the more complicated Rank 3 branch for future work.

	\vspace{-2mm}
	\subsubsection*{Asymptotic charges}
	\vspace{-2mm}
	
	In order to simplify the analysis of the Rank 1 branch, we redefine $k_I$, $\delta$ and $m_3$ in favour of new parameters, $q_I$, $\xi$, $m$, respectively, as follows:
	\begin{gather}
	k_I = \Big(\, 1 + q_1\,,\, 1 + q_2\,,\, 1 + \left( 1 - m \right)\?q_3 \,\Big)\,, 
	\CR
	\delta = \frac1{8\? t_2}\, (a_1 - t_2^2\? a_3)\? \frac{q_1\? q_2\? q_3}{\xi + 1}\,, 
	\CR
	m_I = \Big(\, 0\,,\, 0\,,\, \left( t_2^2\? \frac{a_3}{a_1}  - 1\right)\? m\? q_3 \,\Big)\,.
	\label{eq:delm-def}
	\end{gather}
	Note that in \eqref{eq:delm-def} both the root, $t_2$, and some of the coefficients, $a_i$, of the polynomial $P(u)$ in \eqref{eq:inst-poly} appear, for brevity. One can use \eqref{eq:atr} to write \eqref{eq:delm-def} in terms of the roots $t_i$ only.
	
	With these redefinitions, the conditions on the solution in Section \ref{sec:solution} are summarised as follows. The asymptotic constants are given as in \eqref{eq:AsymPar0},
	while \eqref{eq:as-beh1}--\eqref{eq:as-beh2} are satisfied by fixing the parameter $q_0$ in the harmonic part of the function $M$ in \eqref{eq:M-def} and the parameter, $\gamma$, of the transformation \eqref{eq:t-psi-mix}, as
	\begin{align}\label{eq:AsymPar2}
	q_0 \;=\; &\,{}  - \frac14 + \frac{8\? t_2 }{\kappa}\?\frac1{a_1 - t_2^2\? a_3}\?\frac{\xi^2 - 1}{q_1\? q_2\? q_3}\,,
	\\
	\gamma \;=\; &\,{} -1 - \frac12\? \frac{\xi + 1}{q_1\? q_2\? q_3}\?( q_1\? q_2 + q_1\? q_3 + q_2\? q_3) 
	- \frac{\kappa}{2\? t_2 }\?(a_1 - t_2^2\? a_3)\?\frac{q_1\? q_2\? q_3}{\xi + 1}\,.
	\nonumber
	\end{align}
	The leading terms of the $\mathds{R}^{1,4} \times S^1$ metric at infinity are given by 
	\begin{equation} 
	ds^2 =-dt^2 +  d\yan^2 + \frac{dr^2}{r} +  r \left[  ( d\psi - \cos \theta d\varphi)^2 + d\theta^2 + \sin^2 \theta d\varphi^2 \right] \,,
	\end{equation}
	where the metric of the unit S$^3$ appears in the square bracket.
	For later convenience we parametrize the (standard) lattice of periodicities of these angles by a triplet of free integers ($\ell_1$, $\ell_2$, $\ell_3$) as follows:\footnote{The integer parameters $\ell_1$, $\ell_2$, $\ell_3$ should not be confused with the parameters $l^I$ introduced in~\eq{eq:LI-def} and fixed in \eq{eq:AsymPar0}.}
	\begin{equation} \label{AsymShift} 
	\yan \,\ident\, \yan  + 2\pi \ell_1 R_\yan \; , \quad~~ \psi \,\ident\, \psi + 4\pi \ell_2 - 2\pi \ell_3 \; ,  \quad~~ \varphi \,\ident\, \varphi  +  2\pi \ell_3 \,.
	\end{equation}
It is conventional to take the fundamental domain of this lattice to be given by the ranges $\yan \in [0,2\pi R_\yan),\, \psi\in [ 0,4\pi),\, \varphi \in [0,2\pi)$.
	The asymptotic identifications in \eqref{AsymShift} will be used below to deduce the corresponding periodicities of the compact angles in the interior of the solution.
	
	The explicit expressions for the electric charges $Q_I$ and the parameters $E_I$ appearing in the subleading terms of the $H_I$ are given by\footnote{Note that the parameter $\xi$ corresponds to the parameter $x$ appearing in \cite{Bena:2015drs, Bena:2016dbw}, renamed here to avoid potential confusion with the coordinate $x$ of Section \ref{sec:two-bolt-inst}.}
	\begin{align}
	Q_I \;=\; &\, 4 \frac{ \xi^2-1 }{q_{I+1} q_{I+2}} -  \frac{\kappa}{8\? t_2}\?(a_1 - t_2^2\? a_3)\? q_{I+1} q_{I+2} \,,
	\\
	E_I \;=\; &\, 4\, \frac{ \xi^2-1 }{q_{I+1} q_{I+2}} + \frac{\kappa}{8\? t_2}\?(a_1 - t_2^2\? a_3)\? q_{I+1} q_{I+2} \,,
	\label{eq:EI}
	\end{align}
	which satisfy \eqref{eq:as-beh3}. Similarly, the ADM mass and the two angular momenta defined in \eqref{eq:as-beh2} take the form
	\begin{align}
	M_{\scriptscriptstyle ADM} \;=\; &\, \sum_I E_I \,,
	\CR
	J_\psi \;= \; &\, - \left( 8\,\frac{\xi^2-1}{q_1\? q_2\? q_3} + \frac{\kappa}{4\, t_2}\?(a_1 - t_2^2\? a_3)\?(q_1+q_2+q_3) \right)\,\xi \,,
	\label{eq:ADM} \\
	J_\varphi \;=\; &\, \frac{a_4}{q_1\? q_2\? q_3\,t_2\?(a_1 - t_2^2\? a_3)} \left( \frac1{32}\,j_1\,j_2  - 4\, j_3 \right) ,
	\nonumber
	\end{align}
	where the shorthand combinations $j_1$, $j_2$, $j_3$, read
	\begin{align}
	j_1 \;=\; &\,  t_2^3\, (3\? a_1 - t_2^2\? a_3) + t_1\? t_3\? t_4\? (a_1 - 3\? t_2^2\? a_3) -  2\?  t_2^3\, \, \kappa\, \frac{a_3}{a_4}\, m \,,
	\CR
	j_2 \;=\; &\, \frac{\kappa}{t_2}\, q_1^2\? q_2^2\? q_3^2\, (a_1 - 3\? t_2^2\? a_3) + 32\, ( q_1\? q_2 + q_1\? q_3 + q_2\? q_3)\, (\xi^2 - 1) \,,
	\CR
	j_3 \;= \; &\,t_2^3 \, \kappa\, \frac{a_3}{a_4} \?q_1\? q_2\, m\, (\xi+1)\,.
	\end{align}
	We remind the reader that we use both the $t_i$ and $a_i$ for brevity and that one can use \eqref{eq:atr} to obtain a fully explicit expression.

	\subsection{Structure of the solution around the nuts}
	\label{sec:nuts}
	
	We now turn our attention to the structure of the solution in its interior, where the two three-dimensional cycles are localised. These three-cycles arise from the two-bolt structure of the underlying gravitational instanton, described in Section \ref{sec:two-bolt-inst}. 
	
	The solution topology is clearest in Weyl coordinates $z,\rho$. 
	Consider the spacelike slices defined by fixing $t$ to be constant.
	For $z$ different from the special points $z_i$ \eqref{eq:points}, and different from $z_4=-\infty$, $z_0=\infty$, these spacelike sections are topologically a discrete quotient of $( z_{i+1} , z_{i} ) \times S^1 \times S^1 \times \mathds{R}^2$, where $\mathds{R}^2$ is parametrised in cylindrical coordinates with radius $\rho$. 
	
	In the neighbourhood of $z=z_i$ for $i=1,2,3$, the geometry is a discrete quotient of S$^1 \times \mathds{R}^4$. Recall that we define spherical coordinates $r_i, \theta_i$ centred at $z=z_i,\rho=0$ for each region $( z_{i+1} , z_{i-1} )$, see \eqref{eq:asym-expansions}. 
	The behaviour of the scaling functions as $r_i\rightarrow 0$ is
	\begin{equation} \label{WandHnearcentres} 
	W = \Bigl( e^{-2\,\sigma_i} \frac{N_i}{r_i}\Bigr)^2 + {\cal O}(r_i^{-1})\,, \qquad H_I  =  e^{-2\,\sigma_i} \frac{h_{Ii}}{r_i} + {\cal O}(r_i^{0}) \ , 
	\end{equation}
	for positive constants $N_i,\, h_{Ii}$, where the functions $e^{2\sigma_i}$ were defined in \eqref{sigmaati}. 
	
	In order to avoid closed-time-like curves (CTCs) near the centres, one must impose the behaviour $\mu \approx \frac{1}{r_i}$ near the centres. This behaviour of $\mu$ is ensured provided that the one-form $\omega$ is continuous on the axis; the resulting constraints, known as ``bubble equations'', will be discussed in due course. First however we discuss the geometry near each centre, assuming that these constraints are satisfied. 
	
\begin{figure}[ht]
		\centering
		\includegraphics[scale=.5]{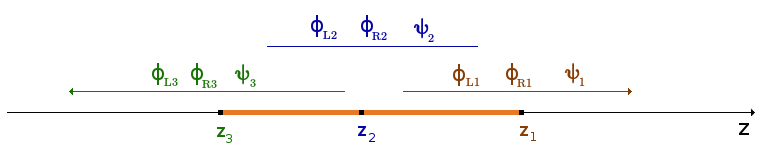}
		\caption{\small The three angular isometries in each of the three coordinate patches around the centres. }
		\label{fig:axis-centres}
	\end{figure}

	To analyze the geometry near the individual centres, we define three different patches with coordinates ($\phi_{Li},\phi_{Ri},\psi_i$), $i=1,2,3$, and which together cover the entire space as depicted in Figure \ref{fig:axis-centres}.
	Specifically, Patch 1 is valid on the open $z$-interval $(z_2, \infty)$; Patch 2 is valid on the open interval $(z_3,z_1)$; and Patch 3 is valid on the open interval $(-\infty,z_2)$.
	
	To explain our notation let us state here which of the angles shrinks on the axis ($\rho=0$) in which region, before deriving this behaviour below.
	Working from right to left, between $z=z_1$ and $z\rightarrow \infty$, $\phi_{R1}$ shrinks at fixed ($\phi_{L1}$, $\psi_{1}$), \ie the norm of the Killing vector $\partial/\partial{\phi_{R1}}$ goes to zero.
	We denote by Bolt A the region between the centres at $z_1$ and $z_2$ on the axis. Here, depending on the patch, it is $\phi_{L1}$ or $\phi_{R2}$ that shrinks.
	Similarly we denote by Bolt B the region between the centres at $z_2$ and $z_3$ on the axis. Here, depending on the patch, it is  $\phi_{L2}$ or $\phi_{R3}$ that shrinks.
	From $z=z_3$ to $z\rightarrow - \infty$, it is $\phi_{L3}$ that shrinks on the axis. 
	Thus our choice of notation should be clear: in Patch $i$, $\phi_{Li}$ shrinks on the left of Centre $i$, and $\phi_{Ri}$ shrinks on the right of Centre $i$, at fixed values of the other coordinates in the patch.

	\vspace{-1.5mm}
	\subsubsection*{Centre 1}
	\vspace{-1.5mm}
	
	One can compute the behaviour of the various functions in the limit $r_1\rightarrow 0$, leading to the following expression for the constant $N_1$ defined in \eqref{WandHnearcentres},
	\begin{equation}\label{eq:N1-def}
	N_1 = \frac{b_\pA+d_\pA}{2}\,,
	\end{equation}
	and the following expressions for the fields $w^0$, $w^3$ and $\alpha^3$ appearing in the metric,
	\begin{eqnarray}
	w^0|_{r_1=0} &=& e^{-2\sigma_1}\Bigl( - \frac{1+\cos \theta_1}{2} + b_\pA d_\pA  \frac{1-\cos \theta_1}{2}\Bigr)\?d\varphi  \,, \CR
	w^3|_{r_1=0} &=& e^{-2\sigma_1} b_\pA k_\pA R_\yan  \frac{1-\cos \theta_1}{2}\?d\varphi  \,, \CR
	\alpha^3|_{r_1=0} &=& - \frac{ k_\pA R_\yan }{b_\pA+d_\pA} \,.
	\end{eqnarray}
	Here $b_\pA$ was defined in \eqref{eq:bAB-def}, and we define the constants $d_\pA$ and $k_\pA R_\yan $ via 
	\begin{align}\label{eq:dA-def}
	d_\pA = &\, - \xi - \frac{\xi + 1 }{(t_1 - t_3)\? (t_1 - t_4)\? (a_1 - t_2^2\? a_3)}\,\left( a_4\? (t_2 - t_1) p_2\? p_3 +  \kappa\? t_2\? (t_1 + t_2)\, \frac{a_3}{a_4}\, m \right) \,, \CR
	\frac{ k_\pA R_\yan}{t_1 - t_2} = &\,
	\left( \frac{\xi^2-1}{\Delta_1\? (a_1 - t_2^2\? a_3)}\?\frac{1}{q_3} 
	- \frac{a_4\?b_\pA}{64}\?  \frac{q_1\? q_2\? q_3}{t_1\? t_2} \right)
	\CR
	& {} ~~~~~ \times 
	\left[ \left( (a_3\?t_1+ a_4\?(t_1 + t_2)^2)\?(t_1 + t_2) - a_1 + t_2^2\? a_3 \right)\,p_1- \frac{a_3}{a_4}\? t_2\? (t_1 + t_2)\? \kappa\, m  \right] 
	\CR
	& {}~~
	-\frac{a_4\?b_\pA}{8}\?  \frac{t_1 + t_2}{t_1\? t_2}\,(q_1+ q_2)\,p_1 
	-\frac{2\,(\xi+1)\,t_2\? (t_1 + t_2)}{\Delta_1\? (a_1 - t_2^2\? a_3)}\,\frac{1}{q_3} \? \frac{a_3}{a_4}\? \kappa\, m \,,
	\end{align}
	where $\Delta_1$ is given by
	\begin{equation} \label{Deltai} 
	\Delta_i = \prod_{\substack{j=1\\ j\ne i}}^{4} (t_i-t_j)  \ .
	\end{equation}

	We wish to impose that the local geometry be smooth up to possible orbifold singularities. By doing so we will explain the physical relevance of the parameters $b_\pA$, $d_\pA$, $k_\pA$ defined above. 
	We consider the metric near $r_1=0$, and we define the Patch 1 coordinates $(\phi_{L1}, \phi_{R1}, \psi_{1})$ via
	\begin{equation} \label{Ch1} 
	\phi_{L1} \,=\, \frac{1}{k_\pA}\?\frac{\yan}{R_\yan}  \ , 
	\qquad~~ \phi_{R1} \,=\,  \varphi +  \frac{b_\pA}{k_\pA}\?\frac{\yan}{R_\yan} \  , 
	\qquad \psi_1 \,=\, \psi - \varphi - 2\? \frac{N_{1}}{k_\pA}\?\frac{\yan}{R_\yan} \,.
	\end{equation}
	Then near $r_1= 0$ the leading spacelike components of the metric (at $dt=0$) are given by (recall that $h_{Ii}$ are defined in \eq{WandHnearcentres})
	\begin{multline}  ds^2 = \frac{h_{31}}{ \sqrt{ h_{11}\?h_{21} }} \Bigl( \frac{ k_\pA R_\yan }{b_\pA + d_\pA} d\psi_1 \Bigr)^2\\
	+ \sqrt{ h_{11}\?h_{21} }\?\Bigl( \frac{dr_1^{\; 2}}{r_1} + r_1 \big(  d\theta_1^{\; 2} +  2(1+\cos \theta_1) d\phi_{L1}^2 + 2 (1-\cos \theta_1) d\phi_{R1}^2\bigr) \Bigr) \,. 
	\end{multline}
	One recognizes the spherically symmetric metric on S$^1 \times \mathds{R}^4$, as anticipated above. The two commuting Killing vectors $\partial_{\phi_{L1}}$ and $\partial_{\phi_{R1}}$ have vanishing norm at $r_1=0$, and they generate rotations of rank four in $\mathds{R}^4$, so we label $r_1=0$ a nut. 
	Let us now analyze the lattice of identifications of the periodic coordinates.
	
	At $\theta_1=\pi$, the direction that shrinks is $\phi_{L1}$ at fixed ($\phi_{R1},\psi_1$). For the shift $\phi_{L1} \to \phi_{L1}+2\pi$ at fixed ($\phi_{R1},\psi_1$) to be a closed orbit, $b_{\pA}$ and $N_1$ must be integers, while for it to be part of the lattice of identifications induced by the identifications of the asymptotic coordinates \eqref{AsymShift}, $k_\pA$ must also be an integer, which we take to be positive without loss of generality.
	
	In full, the identifications of the asymptotic coordinates \eqref{AsymShift} induce the following identifications of the local coordinates \eqref{Ch1}:
	\begin{equation}
	\phi_{L1} \ident  \phi_{L1} + 2\pi \frac{\ell_1}{k_\pA}   \,, 
	\quad~ \phi_{R1} \ident \phi_{R1}  +  2\pi \ell_3+ 2\pi \ell_1 \frac{b_\pA}{k_\pA} \,,  
	\quad~ \psi_1\ident  \psi_1  + 4\pi \left(\ell_2-\ell_3- \ell_1 \frac{N_1}{k_\pA}  \right)   .
	\end{equation}
	Thus the local geometry is a $\mathds{Z}_{k_\pA}$ quotient of  S$^1 \times \mathds{R}^4$; this is a smooth quotient if and only if $N_1$ and $k_\pA$ are relatively prime, when the action of the quotient is free. The geometry is locally identical to that of the orbifolded bolt described in detail in \cite{Chakrabarty:2015foa}, with $m_\pA =N_1$ and $n_\pA = N_1-b_\pA$ the usual integers characteristic of the JMaRT bolt, and $k_\pA$ the order of the orbifold quotient. 
	
	\vspace{-1.5mm}
	\subsubsection*{Centre 3}
	\vspace{-1.5mm}
	
	The structure of the solution near the centre at $r_3=0$ is very similar to the structure at Centre 1, and the analysis is parallel.
	One can again compute the behaviour of the various functions in the limit at $r_3\rightarrow 0$, leading to the following expressions for the constant $N_3$ defined in \eqref{WandHnearcentres}
	\begin{equation}\label{eq:N3-def}
	N_3 = \frac{b_\pB-d_\pB}{2}\,,
	\end{equation}
	and for the metric fields $w^0$, $w^3$ and $\alpha^3$
	\begin{eqnarray}
	w^0|_{r_3=0} &=& e^{-2\sigma_3}\Bigl( b_\pB d_\pB \frac{1+\cos \theta_3}{2} + \frac{1-\cos \theta_3}{2}\Bigr) \?d\varphi \  , \CR
	w^3|_{r_3=0} &=& e^{-2\sigma_3} b_\pB k_\pB R_\yan  \frac{1+\cos \theta_3}{2}\?d\varphi  \  , \CR
	\alpha^3|_{r_3=0} &=&  \frac{ k_\pB R_\yan }{b_\pB-d_\pB} .
	\end{eqnarray}
	Here the quantities $e^{2\sigma_3}$ and $b_\pB$ were defined in \eqref{sigmaati}, \eqref{eq:bAB-def}, and we define the constants $d_\pB$ and $k_\pB R_\yan $ via 
	\begin{align}\label{eq:dB-def}
	d_\pB = &\, - \xi - \frac{\xi + 1 }{(t_1 - t_3)\? (t_3 - t_4)\? (a_1 - t_2^2\? a_3)}\,\left( a_4\? (t_3 - t_2) p_1\? p_2 +  \kappa\? t_2\? (t_3 + t_2)\, \frac{a_3}{a_4}\, m \right) \CR
	\frac{k_\pB R_\yan}{t_3 - t_2} = &\,
	\left( \frac{\xi^2-1}{\Delta_3\? (a_1 - t_2^2\? a_3)}\?\frac{1}{q_3} 
	- \frac{a_4\?b_\pB}{64}\?  \frac{q_1\? q_2\? q_3}{t_2\? t_3} \right)\times
	\CR
	&\,\times 
	\left[ \left( (a_3\?t_3+ a_4\?(t_2 + t_3)^2)\?(t_2 + t_3) - a_1 + t_2^2\? a_3 \right)\,p_3- \frac{a_3}{a_4}\? t_2\? (t_2 + t_3)\? \kappa\, m  \right] 
	\CR
	&\,
	+\frac{a_4\?b_\pB}{8}\?  \frac{t_2 + t_3}{t_2\? t_3}\,(q_1+ q_2)\,p_3 
	+ \frac{2\,(\xi+1)\,t_2\? (t_2 + t_3)}{\Delta_3\? (a_1 - t_2^2\? a_3)}\,\frac{1}{q_3} \? \frac{a_3}{a_4}\? \kappa\, m \,,
	\end{align}
	where $\Delta_3$ is defined in \eqref{Deltai}. 
	
	Near $r_3=0$, we define the Patch 3 coordinates $(\phi_{L3}, \phi_{R3}, \psi_{3})$ via
	\begin{equation} \label{Ch3} 
	\phi_{R3} \,=\, \frac{1}{k_\pB}\?\frac{\yan}{R_\yan} \,, 
	\quad~~ \phi_{L3} \,=\, \varphi + \frac{b_\pB}{k_\pB}\?\frac{\yan}{R_\yan} \,, 
	\quad~~
	\psi_3 \,=\, \psi  + \varphi + 2\? \frac{N_3}{k_\pB}\?\frac{\yan}{R_\yan} \,,
	\end{equation}
	in terms of which the leading spacelike components (at $dt=0$) of the metric near $r_3=0$ are
	\begin{multline}  ds^2 = \frac{h_{33}}{ \sqrt{ h_{13}\?h_{23} }} \Bigl( \frac{ k_\pB R_\yan }{b_\pB - d_\pB} d\psi_3 \Bigr)^2\\
	+ \sqrt{ h_{13}\?h_{23} }\?\Bigl( \frac{dr_3^{\; 2}}{r_3} + r_3 \big(  d\theta_3^{\; 2} +  2(1+\cos \theta_3) d\phi_{L3}^2 + 2 (1-\cos \theta_3) d\phi_{R3}^2\bigr) \Bigr) \,.
	\end{multline}
	In parallel to the analysis near $r_1=0$, we find that $b_\pB$, $N_3$ and $k_\pB$ must be integers. This time we do not have the freedom to chose $k_\pB$ positive (as we did for $k_\pA$), and in all solutions that we have constructed, $k_\pB$ is negative.  We observe that $r_3=0$ is also a nut. 
	
	Under the shift of the asymptotic coordinates \eqref{AsymShift}, one obtains the shift of the local coordinates:
	\eqref{Ch3}
	\begin{equation}
	\psi_3 \,\ident\,  \psi_3  + 4\pi \left(\ell_2+ \ell_1 \frac{N_3}{k_\pB}  \right)   \; , \quad  \phi_{L3}\,\ident\,  \phi_{L3} +2\pi \ell_3 + 2\pi \ell_1 \frac{b_\pB }{k_\pB}   \; ,  \quad \phi_{R3} \,\ident\, \phi_{R3}  +  2\pi  \frac{ \ell_1}{k_\pB} \,,
	\end{equation}
	so that the local geometry is a $\mathds{Z}_{|k_\pB|}$ quotient of  S$^1 \times \mathds{R}^4$ that acts freely if and only if $N_3$ and $k_\pB$ are relatively prime. The geometry is again locally the same as the orbifolded bolt described in detail in \cite{Chakrabarty:2015foa}, with $m_\pB=b_\pB-N_3$ and $n_\pB =-N_3$ the usual integers characteristic of the JMaRT bolt, and $|k_\pB|$ the order of the orbifold quotient.

	\vspace{-1.5mm}
	\subsubsection*{Centre 2}
	\vspace{-1.5mm}
	
	The second centre is the location where the two bolts touch each other, and its analysis is more involved. One can compute the limit of the various functions near $r_2\rightarrow 0$ to obtain the following expressions for the constant $N_2$ in \eqref{WandHnearcentres}:
	\begin{equation}
	N_2 \;=\; \frac{b_\pB d_\pA - b_\pA d_\pB}{2} \;=\; b_\pA N_3 + b_\pB ( N_1-b_\pA )\,,
	\end{equation}
	and for the metric fields $w^0$, $w^3$ and $\alpha^3$:
	\begin{eqnarray}
	w^0|_{r_2=0} &=& e^{-2\sigma_2}\Bigl( b_\pA d_\pA \frac{1+\cos \theta_2}{2} +b_\pB d_\pB \frac{1-\cos \theta_2}{2}\Bigr) \?d\varphi \  , \CR
	w^3|_{r_2=0} &=& e^{-2\sigma_3}  R_\yan  \Bigl( b_\pA k_\pA \frac{1+\cos \theta_2}{2} +b_\pB k_\pB \frac{1-\cos \theta_2}{2} \Bigr) \?d\varphi \  , \CR
	\alpha^3|_{r_2=0} &=&  R_\yan \frac{ b_\pA k_\pB - b_\pB k_\pA  }{b_\pB d_\pA - b_\pA d_\pB} \  , 
	\end{eqnarray}
	where we use the definitions  \eqref{sigmaati}, \eqref{eq:bAB-def} for $e^{2\sigma_2}$ and $b_\pA$, $b_\pB$, and \eqref{eq:dA-def},  \eqref{eq:dB-def} for the constants $d_\pA,d_\pB ,k_\pA ,k_\pB$. 
	
	The Patch 2 coordinates $(\phi_{L2}, \phi_{R2}, \psi_{2})$ are defined by
	\begin{equation} \label{Ch2} 
	\phi_{L2} \,=\, \frac{k_\pA \varphi + b_\pA \tfrac{\yan}{R_\yan}}{k_\pB b_\pA - k_\pA b_\pB} \,, \quad~~~
	\phi_{R2} \,=\, \frac{k_\pB \varphi + b_\pB \tfrac{\yan}{R_\yan}}{k_\pA b_\pB - k_\pB b_\pA} \,, \quad~~~
	\psi_2 \,=\, \psi - d_\pB \phi_{L2} - d_\pA \phi_{R2} \,.
	\end{equation} 
	We find that the leading spacelike components (at $dt=0$) of the metric near $r_2=0$ are 
	\begin{multline}  ds^2 = \frac{h_{32}}{ \sqrt{ h_{12}\?h_{22} } } \Bigl(R_\yan \frac{ b_\pA k_\pB - b_\pB k_\pA  }{b_\pB d_\pA - b_\pA d_\pB} d\psi_2 \Bigr)^2\\
	+ \sqrt{ h_{12}\?h_{22} }\?\Bigl( \frac{dr_2^{\; 2}}{r_2} + r_2 \big(  d\theta_2^{\; 2} +  2(1+\cos \theta_2) d\phi_{L2}^2 + 2 (1-\cos \theta_2) d\phi_{R2}^2\bigr) \Bigr) \,.
	\end{multline}
	Under the shift of the asymptotic coordinates in \eqref{AsymShift} one now obtains 
	\begin{eqnarray}
	(\, \psi_2,\,\phi_{L2},\,\phi_{R2}\,) &\ident&  \left(\, \psi_2+4\pi \ell_2,\,\phi_{L2},\,\phi_{R2}\,\right)
	\CR
	&& + 2\pi \?  \frac{\ell_3}{k_\pB b_\pA-k_\pA b_\pB  }\? \left(\, 2(k_\pA N_3 + k_\pB (N_1-b_\pA) ) ,\, k_\pA ,\, - k_\pB \,\right)
	\CR
	&& + 2\pi\?  \frac{ \ell_1}{k_\pB b_\pA-k_\pA b_\pB}\? \left(\, 2(b_\pA N_3 + b_\pB (N_1-b_\pA) ) ,\, b_\pA ,\, -b_\pB \right) \,.
	\end{eqnarray}
	The local geometry is generally a discrete $\mathds{Z}_{\frac{ |k_{\scalebox{0.4}{$B$}} b_{\scalebox{0.4}{$A$}} - k_{\scalebox{0.4}{$A$}} b_{\scalebox{0.4}{$B$}} |}{{\rm gcd}(b_{\scalebox{0.4}{$A$}},k_{\scalebox{0.4}{$A$}})}} \ltimes \mathds{Z}_{{\rm gcd}(b_\pA,k_\pA)}$ quotient of  S$^1 \times \mathds{R}^4$. To demonstrate this, let us use the existence of the $SL(2,\mathds{Z})$ matrix 
	\begin{equation} \label{gcdrs} 
	g = \left( \begin{array}{cc} \frac{ k_\pA}{{\rm gcd}(b_\pA,k_\pA)} & r_\pA  \\   \frac{ b_\pA}{{\rm gcd}(b_\pA,k_\pA)} & s_\pA \end{array} \right) , 
	\end{equation} 
	for some integers $r_\pA$ and $s_\pA$, to reparametrise the shifts as $( \ell_3^\prime , \ell_1^\prime ) = ( \ell_3 , \ell_1 ) g $, such that the identification reads   
	\begin{eqnarray}\label{PrimeShift}
	(\, \psi_2,\,\phi_{L2},\,\phi_{R2}\,) &\ident&  (\, \psi_2+4\pi \ell_2,\,\phi_{L2},\,\phi_{R2}\,) 
	\CR
	&& + 2\?\pi\? \frac{{\rm gcd}(k_\pA,b_\pA)}{ k_\pB b_\pA-k_\pA b_\pB} \? \ell_3^\prime \,  (\, 2\?N_3 ,\, 1 ,\,0\, ) 
	\CR
	&& + 2\pi \Bigl(  \frac{ \ell^\prime_1}{{\rm gcd}(k_\pA,b_\pA) } + \frac{r_\pA b_\pB - s_\pA k_\pB}{k_\pB b_\pA-k_\pA b_\pB}\? \ell_3^\prime \Bigr)  (\, - 2\? ( N_1 - b_\pA )\? , 0\? ,\, 1\,) \ .
	\end{eqnarray}
	The shift in $\ell_3^\prime$ can be reabsorbed in a shift in $\ell_1^\prime$ for $\ell_3^\prime = 0$ mod ${\frac{ |k_{\scalebox{0.4}{$B$}} b_{\scalebox{0.4}{$A$}} - k_{\scalebox{0.4}{$A$}} b_{\scalebox{0.4}{$B$}} |}{{\rm gcd}(b_{\scalebox{0.4}{$A$}},k_{\scalebox{0.4}{$A$}})}} $, and the shift in $\ell_1^\prime$ is trivial for $\ell_1^\prime = 0$ mod ${\rm gcd}(b_\pA,k_\pA)$. The order of this finite group is therefore $ |k_\pB b_\pA - k_\pA b_\pB |$. The same construction can be done exchanging $A$ and $B$. This group reduces to $\mathds{Z}_{|k_{\scalebox{0.4}{$B$}} b_{\scalebox{0.4}{$A$}} - k_{\scalebox{0.4}{$A$}} b_{\scalebox{0.4}{$B$}} |}$ if either $\gcd (b_\pA,k_\pA)=1$ or $\gcd(b_\pB,k_\pB)=1$. The general condition for the action to be free and the geometry to be smooth is that the shift in $\ell_1$ and $\ell_3$ of the circle coordinate $\psi_2$ must not be an integer multiple of $4\pi$ unless the shifts of both $\phi_{L2}$ and $\phi_{R2}$ are themselves integer multiples of $2\pi$, \ie that any discrete symmetry acting non-trivially on S$^3$ must act non-trivially on S$^1$. This requirement is equivalent to the conditions that
	\begin{eqnarray}
	\label{eq:gcd-conds}
	\frac{  \mbox{gcd}(k_\pA,k_\pB)}{\mbox{gcd}(\,k_\pA N_3 + k_\pB (N_1-b_\pA),\,\,k_\pB b_\pA-k_\pA b_\pB\, )} & \in &  \mathds{Z} \ , \\
	\frac{\mbox{gcd}(b_\pA,b_\pB)  } {\mbox{gcd}(\,b_\pA N_3 + b_\pB (N_1-b_\pA) ,\,\,k_\pB b_\pA-k_\pA b_\pB \,)}&\in &  \mathds{Z}  \ . \nn
	\end{eqnarray}
	These conditions may seem rather difficult to satisfy, but we shall discuss explicit examples in the following. Note that one can reabsorb gcd$(k_\pA,k_\pB)$ in the definition of $R_\yan$, so in practice we work with gcd$(k_\pA,k_\pB)=1$; one can restore gcd$(k_\pA,k_\pB)$ through $R_\yan$ if desired.

	\subsection{Geometry at the two bolts}
	\label{sec:bolts}
	
	We now consider the metric on the two bolts between Centres 1 and 2 and between Centres 2 and 3, denoted as Bolt A and Bolt B respectively. In terms of the spherical coordinates centered at the bolts defined in \eqref{eq:cAB-def}--\eqref{eq:rAB-def}, these correspond to the surfaces $r_\pA=c_\pA$ and  $r_\pB=c_\pB$, respectively. The metric on a spacelike section takes the following form near $r_\pA\rightarrow c_\pA$:
	\begin{multline} \label{eq:6D-metraA}
	ds^2\big|_{\pA} = \,\frac{\hat H^\pA_3}{\sqrt{\hat H^\pA_1 \hat H^\pA_2}}\,
	\left( d\yan +  \ax^3|_\pA \,( d\psi + w^0|_\pA ) + w^3|_\pA  \right)^2 
	\\
	\,        +  \frac{ \hat H_1^\pA  \hat H_2^\pA  \hat H_3^\pA  - \hat \mu^{\; 2}_\pA \sin^2 \theta_\pA}{\hat H_3^\pA \sqrt{\hat H_1^\pA \hat H^\pA_2 } \, \hat W_\pA} \, \sin^2 \theta_\pA \,\scal{  d\psi + w^0|_\pA }^2 \\
	\ + \sqrt{\hat H^\pA_1 \hat H^\pA_2 }\left[ b_\pA^2c_\pA^2\left( \,\frac{dr^2}{r_\pA^2 - c_\pA^2  } + \, d\theta_\pA^2 \right) +  (r_\pA^2-c_\pA^2) \, d\varphi^2 \right] \,,
	\end{multline}
	where $\hat W_\pA$, $\hat \mu_\pA$ and $\hat H_I^\pA$ are regular functions of $\theta_\pA\in [0,\pi]$ that parametrise the values of the corresponding functions at $r_\pA=c_\pA$, as
	\begin{align}\label{eq:bolt-beh}
	W\big|_{r_\pA =c_\pA} = &\, \frac{\hat W_\pA(\theta_\pA)}{\sin^4\theta_\pA} \,, 
	\qquad
	H_I\big|_{r_\pA =c_\pA} =  \frac{\hat H_I^\pA(\theta_\pA)}{\sin^2\theta_\pA} \,,\qquad \mu\big|_{r_\pA =c_\pA} = \frac{\hat \mu_\pA(\theta_\pA)}{\sin^2\theta_\pA} \ .
	\end{align}
	The corresponding expression near the second surface, at $r_\pB=c_\pB$, is obtained from \eqref{eq:6D-metraA}--\eqref{eq:bolt-beh} by replacing $A \rightarrow B$. At these loci, one finds the limiting values 
	\begin{align} &w^0|_{r_\pA=c_\pA} = \frac{d_\pA}{b_\pA} d\varphi \ ,  &w^3|_{r_\pA=c_\pA} = R_\yan \frac{k_\pA}{b_\pA}d\varphi  \ , \CR
	&w^0|_{r_\pB=c_\pB} = \frac{d_\pB}{b_\pB} d\varphi \ ,  &w^3|_{r_\pB=c_\pB} = R_\yan \frac{k_\pB}{b_\pB} d\varphi \ ,
	\end{align} 
	where the integers $b_\pA,\,d_\pA,\, k_\pA$ and $b_\pB,\,d_\pB,\, k_\pB$ are defined as in \eqref{eq:dA-def},  \eqref{eq:dB-def} above in the analysis near the special points $r_i=0$. We can now check that the metric on the spacelike section at the bolt $r_\pA = c_\pA$ is regular in terms of the coordinate systems defined near both the first and the second points, \eqref{Ch1} and \eqref{Ch2}. For this, we use that
	\begin{eqnarray}
	d\yan + w^3|_{r_\pA =c_\pA}  &=& 
	R_\yan \frac{k_\pA}{b_\pA} d\phi_{R1} =  R_\yan \Bigl( k_\pB - \frac{b_\pB}{b_\pA} k_\pA \Bigr) d\phi_{L2} \ , \CR
	d\psi + w^0|_{r_\pA =c_\pA} &=& d\psi_1 + \Bigl( 1 + \frac{d_\pA}{b_\pA} \Bigr)d\phi_{R1} \;=\; d\psi_2 +  \Bigl( d_\pB- \frac{b_\pB}{b_\pA}  d_\pA \Bigr)d\phi_{L2}  \,,
	\end{eqnarray}
	and the fact that the angles whose Killing vectors have zero norm on the bolt, respectively $\phi_{L1}$ and $\phi_{R2}$, only appear in $\varphi$,
	\be
	\varphi \;\simeq\;  - b_\pA \phi_{L1}  \;\simeq\;  - b_\pA \phi_{R2} \,.
	\ee
	One may neglect the other angles in the expression of $\varphi$ because their contribution is subleading in \eqref{eq:6D-metraA}.\footnote{Note that $\phi_{L1} \ne \phi_{R2}$, they are only equal up to corrections in terms of the regular angles at the bolt.} The Killing vector $\partial_{\phi_{L1}}$ defines a rank-two action on the $\mathds{R}^5$ Euclidean tangent space, so we refer to this locus as a bolt.

	In Weyl coordinates, Bolt A is the surface located on the axis $\rho=0$ with $z_2 \le z\le z_1$, and it is described in two patches. We assume that $N_1$ is relatively prime to $k_\pA$. On the first patch, $z\in (z_2,z_1]$; $\phi_{L1}$ is degenerate; $(z,\phi_{R1})$ parametrise a disc $D$ centred at $z=z_1$; and $\psi_1 \in \big[ 0,\frac{4\pi}{{\rm gcd}(k_\pA,b_\pA)} \big)$ parametrises a circle S$^1$, where the periodicity of $\psi_1$ will be explained momentarily. The bolt on this patch is the $\mathds{Z}_{\frac{|k_{\scalebox{0.4}{A}}|}{{\rm gcd}(k_{\scalebox{0.4}{A}},b_{\scalebox{0.4}{A}})}}$ quotient of this $D\times S^1$ by the shift 
	\begin{equation}
	\psi_1 \ident  \psi_1  -4\pi  \ell \? \frac{N_1}{k_\pA}     \; ,  \quad \phi_{R1} \ident \phi_{R1}  + 2\pi \ell \? \frac{b_\pA}{k_\pA}\, .
	\end{equation}
	Note that since we assume that $N_1$ is relatively prime to $k_\pA$, for $\ell=\frac{k_\pA}{{\rm gcd}(k_\pA,b_\pA)}$ this identification means that the periodicity of $\psi_1$ is $\frac{4\pi}{{\rm gcd}(k_\pA,b_\pA)}$, as stated above.

	On the second patch $z\in [z_2,z_1)$, $\phi_{R2}$ is degenerate and  $(z,\phi_{L2})$ parametrise a disc $D$ centred at $z=z_2$. 
	Assuming that $N_1$ is relatively prime to $k_\pA$, the shift in $\ell_1^\prime$ in \eqref{PrimeShift} reduces the periodicity of $\psi_2$ to $\frac{4\pi}{{\rm gcd}(k_\pA,b_\pA)} $, such that $\psi_2 \in \big[0,\frac{4\pi}{{\rm gcd}(k_\pA,b_\pA)} \big)$ parametrises a circle S$^1$.
	Assuming that \eqref{eq:gcd-conds} is satisfied so that the orbifold action is free, 
	the bolt on this patch is then the $\mathds{Z}_{\frac{|k_{\scalebox{0.4}{B}}b_{\scalebox{0.4}{A}}-k_{\scalebox{0.4}{A}} b_{\scalebox{0.4}{B}}|}{{\rm gcd}(k_{\scalebox{0.4}{A}},b_{\scalebox{0.4}{A}})}}$  quotient of this $D\times S^1$ by the shift 
	\begin{equation}
	\psi_2 \ident  \psi_2  + 4\pi  \tfrac{{\rm gcd}(k_\pA,b_\pA) N_3 +(r_\pA b_\pB - s_\pA k_\pB) (b_\pA - N_1)  }{ k_\pB b_\pA-k_\pA b_\pB}  \ell_3^\prime   \; ,   \quad \phi_{L2} \ident  \phi_{L2} + 2\pi  \tfrac{{\rm gcd}(k_\pA,b_\pA)}{ k_\pB b_\pA-k_\pA b_\pB} \ell_3^\prime \; ,
	\end{equation}
	where $r_\pA$ and $s_\pA$ are defined as in \eqref{gcdrs}. 
	The two coordinate sets are related through 
	\be \psi_2 = \psi_2 + 2 \Bigl( \frac{ b_\pB k_\pA - N_1 k_\pB }{k_\pA} - N_3 \Bigr) \phi_{L2} \ , \qquad \phi_{R1} = \frac{k_\pB b_\pA-k_\pA b_\pB}{k_\pA} \phi_{L2} \ . \ee
	In the special case $|k_\pA| = |k_\pB b_\pA-k_\pA b_\pB|$ the entire bolt is a $\mathds{Z}_{\frac{|k_{\scalebox{0.4}{A}}|}{{\rm gcd}(k_{\scalebox{0.4}{A}},b_{\scalebox{0.4}{A}})}}$ quotient of a Hopf fibration over S$^2$, which is a lens space.

	Similarly for Bolt B, the spacelike metric components near $r_\pB = c_\pB$ are regular in terms of the coordinate systems defined near Centres 2 and 3, \eqref{Ch2} and \eqref{Ch3}, and we find
	\begin{eqnarray}
	d\yan + w^3|_{r_\pB = c_\pB}  &=&  R_\yan \Bigl( k_\pA - \frac{b_\pA}{b_\pB} k_\pB \Bigr) d\phi_{R2}  = R_\yan \frac{k_\pB}{b_\pB} d\phi_{L3}   \ , \CR
	d\psi + w^0|_{r_\pB = c_\pB} &=&d\psi_2 + \Bigl(d_\pA - \frac{b_\pA}{b_\pB} d_\pB \Bigr)d\phi_{R2} =d\psi_3 +  \Bigl(  \frac{d_\pB}{b_\pB}-1 \Bigr)d\phi_{L3}  \,,  \CR
	\varphi & \simeq & - b_\pB \phi_{L2} ~\simeq~ - b_\pB \phi_{R3} \,.
	\end{eqnarray}
	In the coordinates around Centre 2, \eqref{Ch2}, the only Killing vector with a vanishing norm on the Bolt A is $ \partial_{\phi_{L2}}$, and the only one on the Bolt B is $ \partial_{\phi_{R2}}$, exhibiting that they are both three-dimensional surfaces in the spacelike section. The topology of Bolt B is similar to that of Bolt A; in particular it is a lens space if $|k_\pB| = |k_\pB b_\pA-k_\pA b_\pB|$.

	\subsection{Absence of closed time-like curves}
	\label{sec:CTCs}
	The final requirement for regularity is the absence of closed time-like curves (CTCs), which would render the solutions pathological. The analysis is identical to the one performed in \cite{Bena:2015drs}, to which we refer for more details; we simply state the relevant conditions here. We start with the determinant of the metric \eqref{eq:6D-metr}, which must never vanish:
	\begin{equation} 
	{} - g \; = \; H_1 H_2 \, e^{4\sigma} \rho^2 \;>\;0 \,.
	\end{equation}
	Combining this with the requirement that the metric component $g_{\yan\yan}$ be positive, and using the base metric \eqref{eq:3d-base}, one finds the conditions
	\begin{equation} \label{eq:reg1}
	H_1 >0 \,, \qquad H_2 >0 \,, \qquad  H_3 > 0  \,.
	\end{equation}
	In addition, demanding that the $g_{\psi\psi}$ and $g_{\varphi\varphi}$ metric components be positive leads to the condition 
	\begin{equation} \label{eq:reg2}
	\frac{H_1H_2H_3-\mu^2}{W} \ge \frac{\omega_\varphi^2}{\rho^2} \,,
	\end{equation}
	which implies that the combination on the left hand side must be everywhere positive and that $\omega_\varphi$ must vanish along the axis of symmetry of the solution, at $\rho=0$. Any regular solution must satisfy both \eqref{eq:reg1}--\eqref{eq:reg2}, however as usual it is not possible to solve these inequalities analytically for the complete family. Instead, we check that these conditions are satisfied on the particular solutions, as discussed in the next section. 
	
	The weaker condition that $\omega_\varphi=0$ along the axis can be written down explicitly, using the analysis near the special points presented above. In particular, the conditions imposed at asymptotic infinity in Section \ref{sec:infinity} ensure that the vector field $\omega_\varphi$ vanishes on the axis $\rho=0$ away from the bolts. On the bolts this is not true automatically and new conditions arise by imposing $\omega_\varphi=0$ on each of the bolts, which (as noted above) also imply that $\mu\approx \frac{1}{r_i}$ at the special points, $r_i=0$. These two conditions have been assumed in deriving some of the above results; we refer to them as ``bubble equations'', in analogy to the similar equations in supersymmetric solutions. 
	Their explicit expressions are given by 
	\begin{align}
	2\,\omega_\varphi\big|_{r_\pA = c_\pA} = &\, 4\,\left( \frac{a_4\?(t_2^2 - t_1^2)}{\kappa\?t_1}\,\frac{p_1}{a_1 - t_2^2\? a_3} + \frac1{4\,b_\pA}\,\frac{d_\pA + 1}{\xi+1}\,\sum_I q_I\? q_{I+1} \right)\,\frac{\xi^2-1}{q_1\? q_2\? q_3}
	\CR
	&\,+\frac{1}{8\?t_2}\,\left( \frac{a_4\?(t_2^2 - t_1^2)}{t_1}\, p_1\,(q_1+q_2+q_3)+ \frac\kappa{4\,b_\pA}\,\frac{d_\pA + 1}{\xi+1}\,(a_1 - t_2^2\? a_3)\,q_1\? q_2\? q_3 \right)
	\CR
	&\, - \left( \frac{t_2 - t_1}{8}\,q_3 + \frac{t_2\,(t_2 + t_1)}{a_1 - t_2^2\? a_3}\,\frac{\xi+1}{q_3} \right)\, \frac{t_2 - t_1}{t_1}\,a_3\,m\,,
	\CR
	2\,\omega_\varphi\big|_{r_\pB = c_\pB} = &\, 4\,\left( \frac{a_4\?(t_3^2 - t_2^2)}{\kappa\?t_3}\,\frac{p_3}{a_1 - t_2^2\? a_3} + \frac1{4\,b_\pB}\,\frac{d_\pB - 1}{\xi+1}\,\sum_I q_I\? q_{I+1} \right)\,\frac{\xi^2-1}{q_1\? q_2\? q_3}
	\CR
	&\,+\frac{1}{8\?t_2}\,\left( \frac{a_4\?(t_3^2 - t_2^2)}{t_3}\, p_3\,(q_1+q_2+q_3)+ \frac\kappa{4\,b_\pB}\,\frac{d_\pB - 1}{\xi+1}\,(a_1 - t_2^2\? a_3)\,q_1\? q_2\? q_3 \right)
	\CR
	&\, - \left( \frac{t_3 - t_2}{8}\,q_3 + \frac{t_2\,(t_3 + t_2)}{a_1 - t_2^2\? a_3}\,\frac{\xi+1}{q_3} \right)\, \frac{t_3 - t_2}{t_3}\,a_3\,m\,,
	\label{eq:bubble-eqns}
	\end{align}
	which must both vanish in order to avoid Dirac--Misner string singularities. We once again remind the reader that both the roots, $t_i$, and the coefficients, $a_i$, of the polynomial parametrising the gravitational instanton appear in this formula for brevity. One can use \eqref{eq:atr}, \eqref{eq:kappa-val} and  \eqref{eq:ni} to express the $a_i$, $\kappa$ and $p_1$, $p_3$ explicitly in terms of the $t_i$.

	Among the eleven independent parameters of this family of solutions, which one can take to be $a_4$, $t_i$, $q_I$, $m$, $\xi$ and $R_\yan$,\footnote{Where we note that $R_\yan$ only arises through the periodicity of the $\yan$ coordinate.} one is redundant and can be fixed by the reparametrisation \eqref{Reparaxy}, and two must be solved for using the bubble equations \eqref{eq:bubble-eqns}. Thus the family of solutions is ultimately parametrised by eight independent parameters, six of which are determined in terms of the six integers $N_1,N_3,b_\pA,b_\pB,k_\pA,k_\pB$ according to the quantization conditions that have been described in this section. The two real parameters, say $q_1,q_2$, and the six integers above are then constrained by the inequalities \eqref{eq:reg1} and \eqref{eq:reg2}. In String Theory, say in the Type IIB D1-D5 frame for concreteness, the D1 and D5 charges as well as the momentum along the $\yan$ circle are quantized such that all the parameters of the solution should eventually be integers, however we will disregard the quantization of charges in this paper.

	\subsection{Two-form potentials and fluxes}
	\label{sec:B-fluxes}
	
	As established in the previous sections, our family of supergravity solutions contains two inequivalent homology three-cycles defined between the three centres, each of which is locally a smooth discrete quotient of S$^1\times \mathds{R}^4$. We can use this fact to compute the fluxes of the three-form field strengths over the two three-cycles, as follows.

	In the adapted coordinates \eqref{eq:Weyl-to-r}, \eqref{Ch1}, \eqref{Ch2}, \eqref{Ch3} around each of the three centres, a given centre is identified as the origin of $\mathds{R}^4$ with the S$^3$ coordinates $\theta_i, \phi_{Li}, \phi_{Ri}$. Since these coordinates are degenerate at $r_i=0$, a regular two-form field must vanish identically along these directions. This implies that a regular two-form with the decomposition \eqref{eq:two-form-exp} can only admit a non-zero component along time and the finite S$^1$ fiber over $\mathds{R}^4$, such that $C_a |_{r_i=0}= C_{ai} dt \wedge d\psi_i$ at  $r_i=0$. We have checked explicitly that the two-form potentials $C_a$ evaluated at each centre $r_i=0$ admit constant components in the base generated by $dt,\, d\yan,\, d\psi,\, d\varphi$ by the wedge product. The corresponding expressions at each centre are rather cumbersome, so we refrain from displaying them. One can remove the unwanted constant components through a distinct gauge transformation on each open set centred at $r_i=0$, with the difference of these constant values determining the gauge transformation from one open set to another, and therefore the flux of the bolt cycle linking them, as we now discuss in some detail.
	
	We define the fluxes, $Q_a^\pA$ and $Q_a^\pB$ on each bolt as the integrals of the field strengths over the surfaces $\Sigma_\pA$, $\Sigma_\pB$ of each cycle, as
	\begin{equation}\label{eq:flux-defs}
	Q_a^\pA \,=\, \frac{1}{4\pi^2}\?\int_{\Sigma_\pA} G_a \,,  \qquad Q_a^\pB \,=\, \frac{1}{4\pi^2}\?\int_{\Sigma_\pB} G_a \, . 
	\end{equation}
	To compute this, we note that the bolt $A$ at $r_\pA =c_\pA$ can be parametrised in the coordinates $z,\rho,\psi_1,\phi_{L1},\phi_{R1}$ defined in \eqref{eq:Weyl-to-r}, \eqref{Ch1} as the surface at $\rho=0$ and at constant $\phi_{L1}$, joining $z=z_1$ to $z=z_2$. The components of the two-forms $C_a$, pulled back to this surface, give
	\begin{equation}
	( C_{a \varphi\psi } d\varphi \wedge d\psi   +C_{a \yan\psi }  d\yan\wedge d\psi  +C_{a\varphi  \yan } d\varphi \wedge  d\yan  )
	\Bigr|_{\phi_{L1}={\rm cst}} \;=\;   C_{a \varphi\psi  }\? d\phi_{R1} \wedge d\psi_1    \ .
	\end{equation}
	Similarly, the bolt $B$ at $r_\pB =c_\pB$ can be parametrised in the coordinates $z,\rho,\psi_3,\phi_{L3},\phi_{R3}$ defined in \eqref{eq:Weyl-to-r}, \eqref{Ch3} as the surface at $\rho=0$ and $\phi_{R3}$ constant, joining $z=z_2$ to $z=z_3$. The components of the two-forms $C_a$, pulled back to this surface, give
	\begin{equation}
	( C_{a \varphi\psi } d\varphi \wedge d\psi   +C_{a \yan\psi }  d\yan\wedge d\psi +C_{a\varphi  \yan } d\varphi \wedge  d\yan )
	\Bigr|_{\phi_{R3}={\rm cst}} \;=\;    C_{a \varphi\psi  }\?  d\phi_{L3} \wedge  d\psi_3 \ .
	\end{equation}
	These coordinates can be used in both cases to compute the flux, since they are well defined everywhere on the two respective bolts, except at the contact point $\rho=0,z=z_2$. One concludes that in both cases one can define the flux integral in the naive coordinates $\theta_\pA , \psi , \varphi$ by taking $\yan$ constant and keeping in mind that the integral is over S$^3/ \mathds{Z}_{|k_\pA|}$,  respectively S$^3/ \mathds{Z}_{|k_\pB|}$.
	Denoting the three patches defined in Section \ref{sec:nuts} by $U_i$ for convenience, the integrals in \eqref{eq:flux-defs} can be re-expressed in terms of the difference of the values of the two-form at the centres as 
	\begin{align}
	Q_a^\pA =&\, \frac{1}{4\pi^2}\,\int_{\Sigma_\pA} d C_a = 
	\frac{1}{4\pi^2}\,\int_{\Sigma_\pA\cap U_1}\!\! d C_a + \frac{1}{4\pi^2}\,\int_{\Sigma_\pA\cap (U_2 \setminus U_1)}\!\!d C_a
	\CR
	= &- \frac{1}{4\pi^2}\,\int_{\Sigma_\pA\cap U_1}\!\!  \left( C_a\Bigr|_{\partial U_1} - C_a\Bigr|_1  \right) + \frac{1}{4\pi^2}\,\int_{\Sigma_\pA\cap U_1}\!\!  \left( C_a\Bigr|_{\partial U_1} - C_a\Bigr|_2 \right) \label{eq:QA} \\
	=&  \frac{2}{|k_\pA|} \left( C_{a\varphi\psi} \Bigr|_1 -  C_{a\varphi\psi} \Bigr|_2 \right) \,,
	\nonumber
	\end{align}
	where the sign comes from the choice of orientation with $z_1> z_2>z_3$. On the second line of \eq{eq:QA}, the second term in each of the two brackets is the distinct gauge transformation performed in the respective patch $U_i$ to set the transformed value of $C$ along the degenerate angles to zero at the special point $r_i=0$.
	Similarly, one obtains
	\begin{equation}\label{eq:QB}
	Q_a^\pB =   \frac{2}{|k_\pB|}\left( C_{a\varphi\psi} \Bigr|_2 -  C_{a\varphi\psi} \Bigr|_3 \right) \; .
	\end{equation}
	The sum of these two fluxes reproduces the total charge, upon taking into account the orbifolding of the bolts by $|k_\pA|$, $|k_\pB|$ mentioned above, as 
	\begin{equation}\label{eq:Q-conserve}
	Q_a =  \frac{1}{4\pi^2}\,\int_{S_\infty^3} G_a = | k_\pA| Q_a^\pA + |k_\pB| Q_a^\pB \,.
	\end{equation}
	
	We emphasise that although the description of Bolt A given above implicitly assumes that the patch $U_1$ is maximal (\ie it only excludes the point $(\rho = 0 , z=z_2)$ in $\Sigma_\pA$), this is by no means necessary. One can check that in general, on a spacelike section $dt=0$, the following equality holds for the two-forms in the patches $U_1$ and $U_2$:
	\begin{equation}
	C_{a} \Bigr|_1 -  C_{a} \Bigr|_2  = \frac{|k_\pA| Q_a^\pA }{2}  d\phi_{R1} \wedge d\psi_1 = \frac{|k_\pA| Q_a^\pA }{2k_\pA} ( k_\pB b_\pA - k_\pA b_\pB )  d\phi_{L2} \wedge d\psi_2 \ . 
	\end{equation}
	Noting that the order of the orbifold action in $U_2$ is $|k_\pB b_\pA - k_\pA b_\pB|$, one obtains the same result for the flux in either coordinate system. Similarly, one finds
	\begin{equation}
	C_{a} \Bigr|_2 -  C_{a} \Bigr|_3  = \frac{|k_\pB| Q_a^\pB }{2}  d\phi_{L3} \wedge d\psi_3 = \frac{|k_\pB| Q_a^\pB }{2k_\pB} ( k_\pA b_\pB - k_\pB b_\pA )  d\phi_{R2} \wedge d\psi_2 \ ,
	\end{equation}
	between the patches $U_2$ and $U_3$. 
	
	Using the solution of Section \ref{sec:solution} in the expression for the two-form potentials in \eqref{eq:two-form-exp} and imposing the regularity constraints analysed in Section \ref{sec:regularity}, one may evaluate \eqref{eq:QA}--\eqref{eq:QB} to find the following explicit expressions:
	\begin{align}\label{eq:fluxes-1}
	|k_\pA| Q_1^\pA = 
	&\, -\frac{1}{\tilde{q}_2} \, \left( 2\,\frac{\xi^2-1}{q_2\?q_3} - \frac{\kappa}{16\,t_2}\, (a_1 - t_2^2\? a_3)\?q_2\?q_3  \right)\,
	\left(1 - \frac\kappa4\, \frac{t_1 + t_2}{t_1 - t_2}\,\frac{q_1\?q_3}{\xi+1}\right)\? q_2\? p_1
	\CR
	&\, -\frac{1}{\tilde{q}_2} \, \left(  \frac{\kappa}{32\?t_2}\?(a_1 - t_2^2\? a_3)\,\frac{q_1\?q_2\?q_3^2}{\xi + 1} + \xi + 1 \right)\, 
	\frac{\kappa}{2}\, t_2\?\frac{a_3}{a_4}\,  q_2\? m \,,
	\CR
	|k_\pB| Q_1^\pB = 
	&\, \frac{1}{\tilde{q}_2} \, \left( 2\,\frac{\xi^2-1}{q_2\?q_3} - \frac{\kappa}{16\,t_2}\, (a_1 - t_2^2\? a_3)\?q_2\?q_3  \right)\,
	\left(1 - \frac\kappa4\, \frac{t_2 + t_3}{t_2 - t_3}\,\frac{q_1\?q_3}{\xi+1}\right)\? q_2\? p_3
	\CR
	&\, +\frac{1}{\tilde{q}_2} \, \left(  \frac{\kappa}{32\?t_2}\?(a_1 - t_2^2\? a_3)\,\frac{q_1\?q_2\?q_3^2}{\xi + 1} + \xi + 1 \right)\, 
	\frac{\kappa}{2}\, t_2\?\frac{a_3}{a_4}\,  q_2\? m \,,
	\end{align}
	where we use the shorthand expression
	\begin{equation}\label{eq:q-tilde}
	\tilde{q}_2 = (t_2 - t_4)\,\left[ (t_1 + t_3)\, q_2 + \frac18\, \frac{t_1 - t_3}{t_2}\, (a_1 - t_2^2\? a_3)\, \frac{q_1\? q_2\? q_3}{\xi+1} \right] .
	\end{equation}
	The corresponding expressions for $Q_2^\pA$ and $Q_2^\pB$ can be obtained from \eqref{eq:fluxes-1}--\eqref{eq:q-tilde} by interchanging the parameters $q_1 \!\? \leftrightarrow q_2$. One can then verify explicitly that \eqref{eq:Q-conserve} is indeed satisfied.

	\newpage
	\section{Exploring the solution space}
	\label{sec:examples-num}
	
	In this section, we summarise the conditions on the parameters of our family of solutions imposed by regularity, and outline a procedure for obtaining solutions to these conditions.
	Using this procedure, we then describe in detail a selection of explicit solutions of interest, representative of our survey of the parameter space.
	
	\subsection{Reparametrisation of variables and summary of constraints}
	\label{sec:summary-constr}
	As briefly explained at the end of Section \ref{sec:CTCs}, the two bubble equations \eqref{eq:bubble-eqns} and the six integrality conditions imposed by smoothness at the three centres in Section \ref{sec:nuts} leave the family of solutions parametrised by these six integers plus two real parameters. Due to algebraic complexity, one cannot parametrise the solution explicitly in terms of these six integers. However one can solve explicitly for four real parameters in terms of 
	$N_1,\, N_3,\, b_\pA,\, b_\pB$, as we shall describe in this section. The only remaining quantization condition to be solved implicitly is the condition that $k_\pA/ k_\pB $ is rational.

	We first recall that the integer quantities $b_\pA$, $d_\pA$, $b_\pB$ and $d_\pB$ depend on the parameters of the solution as in \eqref{eq:bAB-def}, \eqref{eq:dA-def} and \eqref{eq:dB-def}. Starting from the $b$'s, one can solve \eqref{eq:bAB-def} by changing variables to two arbitrary constants $s_1$, $s_2$ as follows, where we immediately fix the unphysical scaling symmetry \eqref{Reparaxy} by setting $s_2=1$ and $s_1=s$:
	\begin{gather}\label{eq:tbAB}
	t_2 = -  \frac{s_1\?s_2\?(b_\pA\?s_1 - b_\pB\?s_2)}{b_\pA\?b_\pB\?(s_1 + s_2)^2 - b_\pA\?s_1^2 - b_\pB\?s_2^2}=-  \frac{s\?(b_\pA\?s - b_\pB\?)}{b_\pA\?b_\pB\?(s + 1)^2 - b_\pA\?s^2 - b_\pB\?}\,,
	\\
	t_1 =t_2+ s_1  = t_2+s \,,
	\qquad
	t_3 =  t_2-s_2  = t_2-1\,,
	\qquad
	t_4 =t_2+  \frac{b_\pA\?s_1 - b_\pB\?s_2}{b_\pA + b_\pB-1}  =t_2+ \frac{b_\pA\?s - b_\pB\?}{b_\pA + b_\pB-1}  \,.
	\nonumber
	\end{gather}
	Similarly, the expressions of $d_\pA$ and $d_\pB$ in \eqref{eq:dA-def} and \eqref{eq:dB-def}  provide a linear system for $\xi$ and $m$ that can also be  solved explicitly. Henceforth, we assume that \eqref{eq:tbAB} is imposed in all relations and that $\xi$ and $m$ are solved for similarly, so that the family of solutions is parametrised by the integers $b_\pA$, $d_\pA$, $b_\pB$ and $d_\pB$ (or equivalently the unconstrained integers $b_\pA$, $b_\pB$ and $N_1$, $N_3$, through \eqref{eq:N1-def} and \eqref{eq:N3-def}), and the real parameters $a_4$, $s$, $q_I$.
	
	We then turn attention to the two bubble equations \eqref{eq:bubble-eqns}, which are both quadratic in each of the three $q_I$, a fact that remains true even after eliminating $\xi$ and $m$ as above, since \eqref{eq:dA-def} and \eqref{eq:dB-def} do not involve the $q_I$. It then follows that one can define a linear combination of the two bubble equations that is linear in $q_3$, which we choose in favour of the others in order to keep manifest covariance under the $GL(1)$ symmetry of the theory in six dimensions. Solving this linear equation for $q_3$, one is left with a single equation which can be seen to be quartic in $a_4$ (\eg by computing the resultant of \eqref{eq:bubble-eqns} in terms of $q_3$ and eliminating $\xi$ and $m$ as explained above). We stress that all operations described up to this point can be done fully analytically and explicitly, since they involve solving at most a quartic polynomial equation; however we refrain from displaying the relevant expressions as they are quite involved and not illuminating.
	
	At this stage, the family of solutions is parametrised by the unconstrained integers $b_\pA$, $b_\pB$ $N_1$, and $N_3$, and the parameters $s$, $q_1$, $q_2$. Additional regularity constraints arise from regularity of the scalar flow near asymptotic infinity, which demands that all $E_I$ in \eqref{eq:EI} are positive, as well as near the three centres, which implies that the functions $H_I$ in \eqref{eq:scal-facts} are positive at all three centres. These properties can be checked explicitly by substituting the results above into the corresponding expressions.
	Finally, one must impose the two constraints required for absence of CTCs, given by \eqref{eq:reg1}--\eqref{eq:reg2}, which are the only conditions that have to be satisfied at all points in the geometry, so that they can only be checked separately for each candidate solution satisfying all other conditions. In practice, we find that these conditions either disallow such a candidate or restrict the allowed ranges of values for the remaining continuous parameters, $s$, $q_1$, $q_2$, for any given set of integers, without fixing them to given values.
	
	As a result, all regular solutions we find are parametrised by the integers $b_\pA$, $b_\pB$ $N_1$, and $N_3$, and the parameters $s$, $q_1$, $q_2$ within allowed ranges. However, one must still impose integrality of the parameters $k_\pA$ and $k_\pB$, which are defined implicitly from their ratio using \eqref{eq:dA-def} and \eqref{eq:dB-def}. In practice one needs to fix $s$ such that $\frac{k_\pA}{k_\pB} \in \mathds{Q}$, which always allows for an infinite number of such choices in any finite allowed range for $s$, although of course one may be less interested in examples where $|k_\pA|$ and $|k_\pB|$ become unreasonably large.

At this point we also make the choice of convention that $t_1>t_2$ which implies $s>0$, so that we focus on the second of the two possible orderings of the roots in \eqref{eq:root-order}.
This is not a requirement for regularity, but rather a conventional choice one can make without loss of generality.

	\subsection{Explicit examples}
	
	In this section, we present a number of interesting explicit example solutions within our family, obtained through the procedure discussed in detail in the previous section. We first discuss a class of solutions featuring an approximate AdS$_3$ region large enough to contain the two-bolt structure, also providing an interesting particular example of such a solution, that in addition has one of the two angular momenta below the regularity bound. We then turn to a sub-family of solutions that allows for a parametrically small charge-to-mass ratio, describing some of its salient features.
	We also comment on the properties of a multitude of other solutions with an intermediate amount of non-extremality with respect to charge, that we obtained through computer-aided scans of the parameter space of smooth solutions, however we will not discuss these solutions in detail.
	
	\vspace{-1.5mm}
	\subsubsection{Solutions featuring an AdS$_3$ region}
	\label{sec:AdS3}
	\vspace{-1.5mm}
	
	It is interesting to consider the near-supersymmetric regime, and to investigate whether our family of solutions contains examples with an approximate AdS$_3$ region. This would suggest the possibility of taking a decoupling limit and investigating the resulting solutions holographically. The JMaRT solutions \cite{Jejjala:2005yu} can be studied in the near-supersymmetric limit, which allows for solutions exhibiting an AdS$_3$ near-core region. The decoupled JMaRT solutions have a now well-understood holographic interpretation \cite{Chowdhury:2007jx,Chakrabarty:2015foa}. 
	
	While the complexity of our family of solutions does not allow for a general direct analysis, it indeed contains solutions with an approximately AdS$_3\times$S$^3$ region. It would be very interesting to investigate whether one can take a decoupling limit of these solutions, whether they can be studied holographically, and whether one can ultimately connect to recent holographic studies of black hole formation in the D1-D5 system~\cite{Carson:2014yxa,Carson:2014xwa,Carson:2014ena,Carson:2015ohj,Carson:2016cjj,Carson:2016uwf}.
	
	Performing a computer-aided analysis along the lines of Section \ref{sec:summary-constr}, one can identify a plethora of solutions with an approximate AdS$_3\times$S$^3$ region, characterised as 
	\begin{equation}
	\frac{E_1}{E_3}  \gg 1\,, \qquad \frac{E_2}{E_1} = {\cal O}(1)\,, \qquad \frac{E_1}{|\xi_\infty|} \gg 1\,,
	\end{equation}
	where $\xi_\infty$ was defined in \eqref{eq:as-beh1}.
	These conditions imply that there exists a region $E_1, E_2 \gg r \gg E_3 , |\xi_\infty| $ where the scaling functions are well approximated by the AdS$_3\times$S$^3$ solution
	\begin{equation}
	H_1 \approx \frac{E_1}{4r^2} \ , \quad H_2 \approx \frac{E_2}{4r^2} \  , \quad H_3 \approx \frac{1}{r}  \ , \quad W \approx \frac{1}{r^2} \ .
	\end{equation}
	The solutions of this type are very similar to the JMaRT solutions at large distances and up to the five-dimensional ergoregion (the ergoregion of the solution obtained by dimensional reduction on $\yan$), where $r\approx E_3>4 |\xi_\infty|$ and $W$ changes sign, but $r\gg c_\pA , c_\pB$. In this intermediate region the scaling factors behave similarly as for  AdS$_2\times $S$^4$,
	\begin{equation}
	H_1 \approx \frac{E_1}{4r^2} \ , \quad H_2 \approx \frac{E_2}{4r^2} \  , \quad H_3 \approx \frac{E_3+4 \xi_\infty \cos \theta}{4r^2} \ ,
	\end{equation}
	however the geometry is strongly modified by the presence of the ergoregion because $|\xi_\infty|$ is not small compared to $E_3$. This region is itself widely separated from the neighbourhood of the two bolts, {\it i.e.}~where $\frac{r_\pA}{c_\pA}$, $\frac{r_\pB}{c_\pA} \sim {\cal O}(1)$. We thus find three separations of scale in such solutions, where the parameters are such that $(E_1,E_2)  \gg ( E_3 , |\xi_\infty|) \gg ( c_\pA , c_\pB)$. As one moves from asymptotic infinity towards the centre of the geometry, the different regions are summarised as follows:
	\begin{align}\label{eq:AdS3-regions}
	\mathds{R}^{1,4}\times \mbox{S}^1\,, \,\quad  &\,r\gg E_1 \,,
	\CR
	\mbox{AdS}_3\times \mbox{S}^3 \,, \,\quad  &\, E_1 \gg r \gg E_3  \,,
	\CR
	\mbox{5D}\text{ ergoregion} \,, \,\quad  &\, E_3 \gg r \gg c_\pA \,,
	\CR
	\text{two-bolt region} \,, \,\quad &\, E_3 \gg  r \approx 2 (c_\pA+c_\pB) \,,
	\end{align}
	Such asymptotically-flat examples featuring long throats well-approximated by AdS$_3\times$S$^3$ hint towards the existence of a decoupling limit, to which we hope to return in future work.
	
	We now discuss in more detail an explicit example, setting for simplicity
	\begin{equation}\label{eq:AdS3-pars}
	q_1=q_2= c^{-1/2}\,,
	\end{equation}
	where $c$ is a real positive constant that serves as an arbitrary length scale. This implies that the two electric charges are equal, $Q_2=Q_1$, and also implies that $E_2=E_1$. Following the procedure described in the previous section, we start by choosing a set of integers parametrising the geometry at the three centres, as 
	\begin{equation}\label{eq:fav-AdS3}
	b_\pA = 23,, \quad b_\pB = -158\,, \quad N_1 = -9\,, \quad N_2 = - 50\,, \quad N_3 = - 222\,.
	\end{equation}
	Using the change of variables \eqref{eq:tbAB} along with \eqref{eq:AdS3-pars} in the bubble equations \eqref{eq:bubble-eqns}, one can find a solution for the parameters $a_4$ and $q_3$ that depends on the single remaining free parameter, $s$, as described in the preceding section. This free parameter can be used to tune the ratio $k_\pA/k_\pB$ defined by \eqref{eq:dA-def} and \eqref{eq:dB-def} to be a rational number, such that the conditions described in Section \ref{sec:nuts} for the absence of orbifold singularities at the special points $r_i=0$ are satisfied. One such choice, that in addition allows for a good approximate AdS$_3$ region, is given by
	\begin{gather}
	s \approx 4.01531 \quad\Rightarrow\quad k_\pA= 133\,, \quad k_\pB =-935\,,
	\CR
	a_4\approx -3.00589\? c\,, \quad q_3\approx 0.000988359 \? c^{-1/2}\,, \label{eq:kAkB-AdS-example}
	\end{gather}
	where we also give the resulting value for the constrained parameters $a_4$ and $q_3$. We observe that $\gcd(N_1,k_\pA)=1$ and $\gcd(N_3,k_\pB)=1$, and that the two ratios in \eqref{eq:gcd-conds} are both equal to one. One can further check that all the relevant quantities \eqref{eq:reg1} and \eqref{eq:reg2} are positive everywhere; thus the solution describes a globally hyperbolic smooth manifold.

	With the numerical values above, one can now directly evaluate all physical quantities for this example solution. Starting from the asymptotic charges, we find that 
	\begin{gather}
	E_1 = E_2 \approx 8.35107\times10^8\? c\,,\quad E_3 \approx 1.07441\times10^6\? c\,,
	\CR
	Q_1 = Q_2 \approx 8.35106\times10^8\? c\,,\quad Q_3 \approx 5.76365\times10^5\? c\,.
	\end{gather}
	The resulting ratio $E_1/E_3 \approx 780$, together with $\xi_\infty \approx -2.26678 \times10^5\? c$ (with $4|\xi_\infty| < E_3$), signals a good approximate AdS$_3$ region as anticipated above, and thus a near-extremal solution. This can be verified by using \eqref{eq:ADM} to compute the ratio of the ADM mass to the BPS mass, 
	\begin{equation}
	\frac{M}{\sum_I |Q_I|}\approx1.0003 \,.
	\end{equation}

\begin{figure}[h!]
		\centering
		\includegraphics[scale=.15]{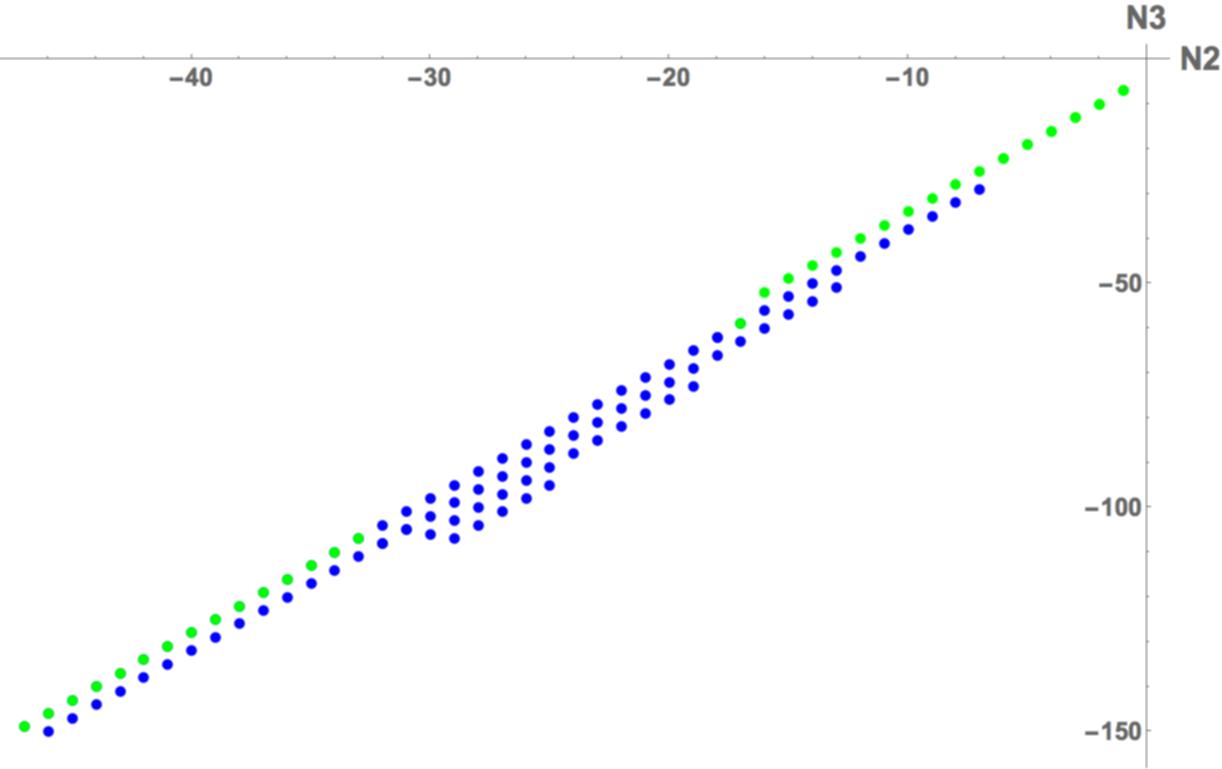}
		\hspace{.5cm}
		\includegraphics[scale=.166]{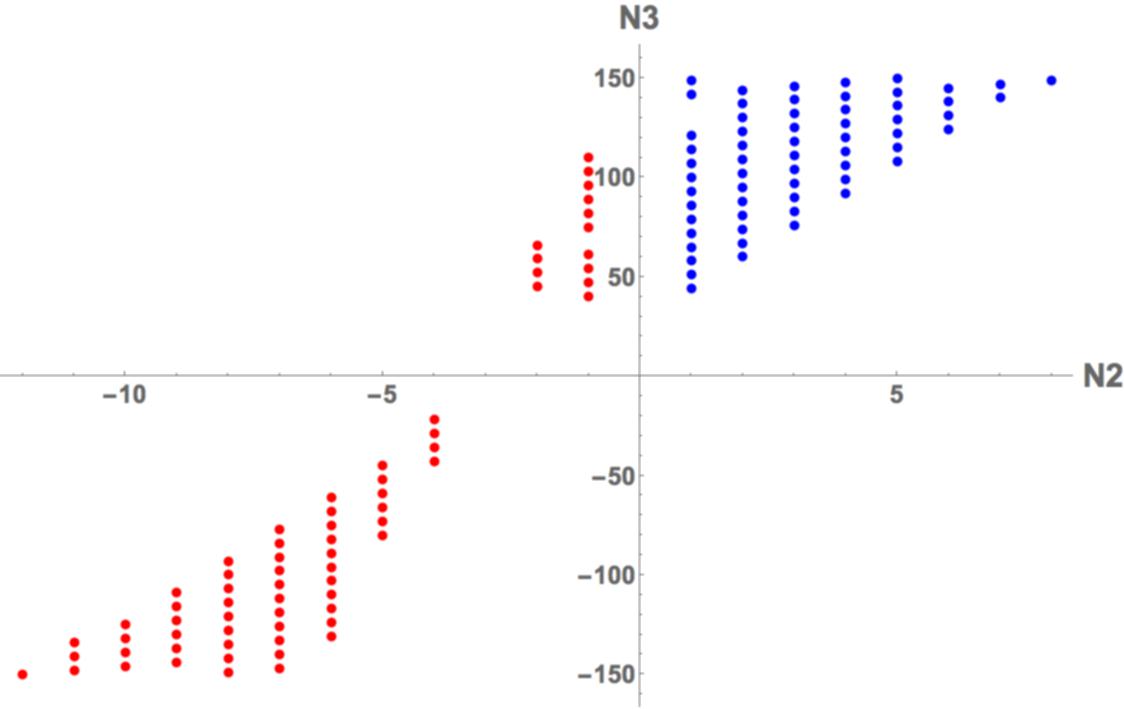}
		\caption[caption]{\small Examples of integer parameters for various solutions. \\\hspace{\textwidth}
			{\it Left}: Solutions obtained setting $(N_1,\, b_\pA)=(-1,\,3)$, while scanning for $(N_2,\, N_3)$. All solutions have opposite signs of fluxes on the two bolts. The blue points represent solutions breaking the regularity bound for both angular momenta, while the green points represent solutions where only the bound for $J_\psi$ is broken. \\\hspace{\textwidth}
			{\it Right}: Solutions obtained setting $(N_1,\, m_\pB)=(-3,\,7)$, while scanning for $(N_2,\, N_3)$. All solutions break both regularity bounds for the angular momenta, while the blue (red) points represent solutions with the same (opposite) sign of fluxes on the two bolts.}
		\label{fig:Sol-families}
	\end{figure}

	Similarly, the angular momenta can be also computed from \eqref{eq:ADM}, in order to compare with the relevant regularity  bounds for a non-extremal black hole \cite{Cvetic:1996kv,Cvetic:2011hp}, which read
	\begin{align}\label{eq:J-bounds}
	|J_\psi| < &\, J_{\text{max}}^+  =  \tfrac{1}{2\?\sqrt{2}}\? \Bigl({ \scriptstyle \sqrt{ (E_1 +Q_1)(E_2+Q_2)(E_3+Q_3)} \;+\;   \sum\limits_I \sqrt{ (E_I +Q_I)  (E_{I+1} -Q_{I+1}) (E_{I+2} -Q_{I+2})}} \Bigr) \ , 
	\CR 
	|J_\varphi| < &\, J_{\text{max}}^-  =  \tfrac{1}{2\?\sqrt{2}}\? \Bigl({ \scriptstyle \sqrt{ (E_1 -Q_1)(E_2-Q_2)(E_3-Q_3)} \;+\;   \sum\limits_I \sqrt{ (E_I -Q_I)  (E_{I+1} +Q_{I+1}) (E_{I+2} +Q_{I+2})}} \Bigr) \ . 
	\end{align}
	For the explicit solution above, we find
	\begin{equation}
	\frac{|J_\psi|}{J_{\text{max}}^+}  \approx 1+2.42311\times10^{-6} \,, 
	\qquad 
	\frac{|J_\varphi|}{J_{\text {max}}^-}   \approx 1 - 2.20732\times10^{-5} \,,
	\end{equation}
	so that both angular momenta are close to the bound, but only one satisfies it. This implies that this solution does not have the asymptotics of a regular five-dimensional black hole, but corresponds to an over-rotating solution along $\psi$. This is the case for all solutions considered in this family, which have at least the angular momentum $J_\psi$ violating the bound, regardless of whether they contain an approximate AdS$_3$ region or not, see \eg the left panel on Fig. \ref{fig:Sol-families}.

	Moving deeper into the bulk, we find that the radii of the bolts are order one in units of $c$,
	\begin{equation}
	c_\pA \approx 4.40073\? c\ , \qquad c_\pB \approx  2.5371\? c\ ,
	\end{equation}
	and are therefore small compared to the asymptotic circle radius $R_\yan \approx 140908\? c^{1/2}$ at infinity. We also record the values of the fluxes on the two bolts, as computed by \eqref{eq:fluxes-1}, given by
	\begin{equation}
	Q_1^\pA = Q_2^\pA \approx -1.98574\times10^6\? c\,,\quad Q_1^\pB = Q_2^\pB \approx 1.17563\times10^6\? c\,,
	\end{equation}
	where the two fluxes on each bolt are equal due to the choice in \eqref{eq:AdS3-pars}. Note that the signs of the fluxes on the two bolts are opposite: this may suggest a microscopic origin as a bound state of D1/D5 branes and D1/D5 anti-branes. The topology of the solution is such that the total flux at infinity is much larger than the sum of the fluxes on the two bolt cycles, because of the fairly large values of $k_\pA$ and $|k_\pB|$ in \eq{eq:kAkB-AdS-example}, that feed into the expression for global charges \eqref{eq:Q-conserve}. The lens-space nature of the two bolt cycles as $\mathds{Z}_{n_1} \times \mathds{Z}_{n_2} $ quotients of S$^3$ with a large order $n_1 n_2$ generates a kind of ``gravitational lens'' effect,\footnote{Here we do not mean gravitational lensing, in the sense of the bending of light, but rather we are trying to convey the amplification of the interior fluxes by the topology of the solutions.} that amplifies the interior fluxes to produce a rather large total charge compared to the total flux in the deep  interior.

	More generally, our investigation suggests that it is difficult to find globally smooth solutions in this family that have no orbifold singularities, and that have an approximate AdS$_3\times $S$^3$ throat, without having somewhat large $k_\pA$ and $|k_\pB|$.
	We do not presently have a general understanding of this feature, due to the necessity of relying on explicit examples with particular values of the parameters.  However 
	the pattern appears to be that the condition $E_1 \gg E_3$ is only valid on a very small interval for the parameter $s$, for which the value of $\frac{k_\pA}{k_\pB}$ can only vary by a very small amount. With such a constraint, the available rational numbers naturally have rather large numerators and divisors. 
	
	Our family of solutions exhibits the feature of carrying fluxes of opposite sign on the two bolt cycles for various choices of integer parameters,  not restricted to solutions containing an AdS$_3$ region, see Fig.\;\ref{fig:Sol-families}. It would be very interesting to investigate whether these flux/anti-flux topological structures can be interpreted as brane/anti-brane bound states dissolved in flux within string theory, and to investigate the stability properties of such configurations.

	\vspace{-1.5mm}
	\subsubsection{Solutions with small charge-to-mass ratio}
	\label{sec:large-M}
	\vspace{-1.5mm}
	
	We now turn to a discussion of a two-parameter sub-family of solutions, that allows for a parametrically small charge-to-mass ratio, \ie for asymptotics deep in the non-extremal region with respect to charge. 
	This is, in a sense, the opposite regime to the near-supersymmetric regime discussed in the previous paragraph: there the mass above extremality was small, here it will be large.
	
	We consider the following set of integers,
	\begin{equation}\label{eq:large-M-integers}
	b_\pA = 1 + N_1\,, \quad b_\pB = - 1 -p\? N_1\,, \quad N_2 = (p-1)\?N_1\,, \quad N_3 = - 1\,,
	\end{equation}
	where $p$ is a rational parameter such that $p N_1 \in \mathds{Z}$. 
	We further fix $q_1=q_2 = c^{-\frac{1}{2}}$ as in \eqref{eq:AdS3-pars}, for simplicity. One can then solve the constraints required by regularity of the solution, as summarised in Section \ref{sec:summary-constr}, order-by-order in an expansion in $N_1^{-1/4}$. It is straightforward to verify that this expansion is regular for any value $1< p < \frac{3+\sqrt{5}}{2}$, but the expressions involved are rather long and not illuminating, so we henceforth concentrate on the case $p=2$ (so that $N_1$ is an arbitrary large integer), and give the various quantities up to subleading corrections in $N_1^{-1/4}$.
	
	With the values of parameters above, we perform the change of variable
	\begin{equation}
	s = \frac{1+\sqrt{5}}{2} + \frac{\sigma}{N_1^{1/4}} \,,
	\end{equation}
	where $\sigma>0$ is a real parameter that is taken to be of order 1 with $N_1$, whereas $N_1^{1/4}\gg 1$. Using this, we solve the bubble equations for the scale invariant quantities $a_4/c$ and $q_3\?c^{1/2}$, to find the asymptotic expansions
	\begin{align}
	\frac{a_4}{c} =&\, 4\?(-3 + \sqrt{5})\?\left( 1- \frac{4\?\sigma}{N_1^{1/4}} \ \right) + \frac{4}{ N^{\frac{1}{2}}} \Bigl( - 4 \frac{ 3+ \sqrt{5}}{5\sigma^2} + ( 11\sqrt{5} -39) \sigma^2 \Bigr) + {\cal O}(N_1^{-3/4})\,,
	\CR
	q_3\?c^{1/2} = &\, 1 + \frac{\sqrt{5}}{2}\?(3 - \sqrt{5})\?\frac{\sigma}{N_1^{1/4}} + \frac{1}{ N^{\frac{1}{2}}} \Bigl( -\frac{ 7+ 3\sqrt{5}}{5\sigma^2} + \frac{3}{8} ( 19-9 \sqrt{5}) \sigma^2 \Bigr) +  {\cal O}(N_1^{-3/4})\,,
	\end{align}
	which allow us to compute all relevant quantities at leading order in the large $N_1$ expansion. One can check in the asymptotic expansion that the poles of the $H_I$ functions are strictly positive at large $N_1$. One can moreover check all the regularity bounds for a series of explicit examples with a fixed numerical value of $N_1$, to find that they are indeed satisfied. 
	The angular momenta reach the extremality bound from above at large $N_1$, with a strictly positive $\mathcal{O}(N_1^{-1/4})$ correction.
	
	The mass to BPS bound ratio is given by:
	\begin{equation}
	\frac{M}{\sum_I |Q_I|} = \frac{3}{20}\?(5 + 3\?\sqrt{5})\?\frac{N_1^{1/4}}{\sigma} + \frac{3}{80}\?(23 + 9\?\sqrt{5}) - \frac{3}{25}\?(47 + 21\?\sqrt{5})\?\frac{1}{\sigma^4}  + {\cal O}(N_1^{-1/4})\,, 
	\end{equation}
	which becomes arbitrarily large as one increases $N_1$. One may use the scale invariance of the system expressed through the arbitrariness of the scale, $c$, to rescale the mass to a finite value at large $N_1$, by performing a change of variable $c\rightarrow N_1^{-2} c$. The resulting electric charges vanish as $Q_1=Q_2 = \mathcal{O}(N^{-1/4})$ and $Q_3 =  \mathcal{O}(N^{-1/2})$, while the ratio of the fluxes on the two bolts behaves as
	\begin{equation}
	\frac{Q_1^\pA}{Q_1^\pB} = \frac{5-2\?\sqrt{5}}{N_1^{1/4}}\?\sigma\? + \mathcal{O}(N^{-1/2})\,,
	\end{equation} 
	so that the flux on Bolt B dominates at large $N_1$. While the flux on Bolt A is subleading in the large $N_1$ limit, this bolt is not completely irrelevant, as the geometry near the remaining point away from Bolt B, at $r_1=0$, is very intricate at large $N_1$.

	The behaviour of $R_\yan$ at large $N_1$, after the rescaling defined above, is $R_\yan = \mathcal{O}(N_1^{\frac12})$,  so this is a large $R_\yan$ limit at finite mass. At this point one might imagine that the supergravity approximation is good everywhere, however it is straightforward to compute that
	\begin{equation}
	\frac{k_\pA}{k_\pB} = - \frac{1}{2} + \frac{ 2 + \sqrt{5} }{ 5\? \sigma^2}\?N_1^{-1/2} + \mathcal{O}(N_1^{-3/4}) \ , 
	\end{equation}
	so that $k_\pA$ and $|k_\pB|$ are necessarily very large positive integers, of the order of $N_1^{1/2}$, so one cannot trust the supergravity approximation near the bolts for arbitrarily large $N_1$. Moreover, the rescaling  to finite mass implies that the characteristic scales of the solution are arbitrarily small; for example this is in contradiction with the quantisation of the flux $Q_1^\pA \sim N_1^{-1}$. In order to be compatible with flux quantisation, one should not perform the above rescaling, and instead keep $N_1$ large but finite.
	
	\vspace{-1.5mm}
	\subsection{General remarks on the solution space}
	\vspace{-1.5mm}
	
	We conclude this section with some general remarks on the properties of solutions obtained by exploring the bubble equations for various sets of integers. It appears to be a general feature that all solutions in the family constructed in this paper have at least one of the angular momenta larger than the black hole regularity bound, \ie $J_\psi> J_{\text{max}}^+$, whereas $J_\varphi$ may either satisfy the regularity bound \eqref{eq:J-bounds} or not. This is similar to the solutions of \cite{Bena:2015drs}, which also contain three centres. 
	
	We have identified solutions with both aligned and anti-aligned signs of fluxes, for various regions of the parameter space. 
	We find that the examples featuring the best approximate AdS$_3$ throats---including the one presented in Section \ref{sec:AdS3}---arise for solutions with opposite signs of fluxes on the bolts. 
	By contrast, the far-from-extremal solutions we have obtained, including the family in Section \ref{sec:large-M}, all have fluxes of the same sign on the two bolts. 
	This seems both intriguing and somewhat counter-intuitive, since one might have expected a priori that solutions featuring fluxes of opposite signs are more likely to correspond to the far-from-extremal regime with respect to the overall charges. 

	Note that these observations are heavily influenced by our incomplete understanding of the solution space; thus we would caution against concluding at this point that there must be a physical reason for this apparent correlation between degree of non-extremality and the relative sign of the fluxes on the two bolts (although if such a reason exists, it would of course be interesting to elucidate it). The family of Section \ref{sec:large-M} was explicitly motivated by its similarity to the single-bolt solution, thus allowing for an expansion in a large parameter and a simplification of the bubble equations. In principle, our scan for solutions featuring an approximate AdS$_3$ throat is more systematic, however the parametrisation of our family of solutions is in terms of parameters that are not directly physical, so it is conceivable that a significant component of the allowed solution space may reside in a small corner of this parameter space that lies outside our scan. 
	
	To conclude, we have observed a remarkably rich amount of physics in this family of two-bolt supergravity solutions. There is much scope for further study, most obviously the construction of more general classes of supergravity solutions building upon our results. We hope to investigate these new avenues in future work.

	\vspace{2mm}
	
	\section*{Acknowledgements}
	
	The authors thank A. Virmani for bringing to their attention the two-bolt instanton solution of \cite{Chen:2011tc,Chen:2015vva}, and thank A. Virmani, N. Warner, and especially I. Bena for stimulating discussions. The work of GB and DT was supported by the ANR grant Black-dS-String. The work of SK is supported in part by the KU Leuven C1 grant ZKD1118 C16/16/005, by the Belgian Federal Science Policy Office through the Inter-University Attraction Pole P7/37, and by the COST Action MP1210 The String Theory Universe. The work of DT was supported by a CEA Enhanced Eurotalents Fellowship and by a Royal Society Tata University Research Fellowship.

	\newpage
	\begin{appendix}
		
		\section{Explicit expressions for the supergravity ansatz}
		\label{app:system}
		In this appendix we collect the relevant explicit expressions for the various fields appearing in the supergravity ansatz of Section \ref{sec:sugra-sol}, as given in \cite{Bena:2016dbw} and used throughout this paper.
		Starting from the vector fields, the expressions for $\omega$, $w^0$ and $w^I$ are determined by the first-order equations
		\begin{align} \label{wIEq-1} 
		\star d \omega =&\, d M - \frac{V}{1+V\Vb}\, \left( \sum^3_{I=1}L^I \,d K_I -2\, M \, d \Vb \right) \,,
		\CR
		\star d w^0 =&\, -(1+\Vb)\, d M -\frac{1}{2}\,\frac{1-V\Vb-2\,V}{1+V\Vb}\, \left( \sum^3_{I=1}L^I \,d K_I -2\, M \, d \Vb \right) 
		+ \frac{1}{2}\,\sum^3_{I=1} K_I \,d L^I
		\CR
		&\,   - \frac1{4}\,\frac{V}{1+V\Vb}\,d\left( K_1\? K_2\? K_3 \right)
		+ \frac1{4}\,\frac{K_1\? K_2\? K_3}{(1+V\Vb)^2}\left( V^2 d\Vb + dV \right) \,, 
		\\
		\star d w^I =&\, d L^I - \frac{1}{2}\, \frac{V}{1+V\Vb}\,d\left( K_{I+1} K_{I+2} \right)
		+ \frac{1}{2\,(1+V\Vb)^2}\,K_{I+1} K_{I+2} \left( V^2 d\Vb + dV \right) \, ,
		\nonumber
		\end{align}
		where the triplet $w^I$ contains both the vector field $w^3$ appearing in the metric through \eqref{eq:6d-KK} and the $w^a$ appearing in the two-form potentials in \eqref{eq:two-form-exp}.
		
		Similarly, the one-forms, $v_a$, $b_a$ in \eqref{eq:two-form-exp} are determined in terms of the functions appearing in the ansatz  by solving the first-order equations
		\begin{align}
		\label{baEq-2} 
		\star db_a ~=~&\, \frac{1-V}{1+V\Vb}\,d K_a + \frac{K_a}{(1+V\Vb)^2}\,\left((V-1)\, V\, d \Vb + (1 + \Vb) d V \right)\,,
		\\
		\label{eq:alm-NE-mag}
		\star d v_a ~=~ &\, -\frac{V}{1+V\Vb}\,d K_a + \frac{K_a}{(1+V\Vb)^2}\left( V^2 d\Vb + dV \right)
		\,.
		\end{align}
		
		Finally, the electric potentials $A_t^a$, the axions $\ax^a$ and the scalars $\beta_a$ in \eqref{eq:two-form-exp}, along with the scalars $A_t^3$, $\ax^3$ of \eqref{eq:6d-KK} are given by
		\begin{align}
		A^I_t = & \,\frac1{2\,H_I}\left( 2\, (1 + \Vb)\, M - \sum^3_{J=1} K_J L^J 
		+\frac{1}{2} \frac{V\,K_1 K_2 K_3}{1+V\Vb} - 2\, \frac{V - 1}{1 + V\Vb}\,K_I L^I \right)\,,
		\label{eq:5dzeta}
		\\
		\ax^I = &\, \frac{1}{H_I}\left( M - \frac{V\, K_I L^I}{1+V\Vb}\right) \,,
		\label{eq:5dax}
		\\
		\beta_a = & \, \frac{H_a}{H_1 H_2}\left( L^3  - \frac{V}{1 + V\,\Vb}\? K_1 K_2 \right) \, .
		\end{align}
		where we again use a triplet notation to group these quantities where possible.

		\newpage
		\section{Adapted coordinates and expansions at the special points}
		\label{app:coord-change}
		
		In this appendix we give some more details on the coordinate changes in \eqref{eq:Weyl-to-r}, which are useful in the expansions around the various interesting points of the solution. We start by writing down the explicit inverse of $r_i$ as functions of $x$ and $y$ given by \eqref{eq:Weyl-to-r} together with \eqref{eq:Weyl-coord}; this takes the simple form
		\begin{equation}\label{eq:xyTri}
		x = \frac{2\,(n_1\?r_1 + n_2\?r_2 + n_3\?r_3)  + 1}{2\,(f_1\?r_1 + f_2\?r_2 + f_3\?r_3) } \,,
		\qquad
		y = \frac{2\,(n_1\?r_1 + n_2\?r_2 + n_3\?r_3)  - 1}{2\,(f_1\?r_1 + f_2\?r_2 + f_3\?r_3) } \,,
		\end{equation}
		where the constants $n_i$ and $f_i$ are given by 
		\begin{align}
		n_1 = &\, \frac{a_4}{4}\,\frac{t_3\? t_4 - t_1\? t_2}{(z_2 - z_1)\? (z_1 - z_3)}\,,
		\qquad
		f_1 = \frac{a_4}{4}\,\frac{t_3 + t_4 - t_1 - t_2}{(z_2 - z_1)\? (z_1 - z_3)}\,,
		\CR
		n_2 = &\, \frac{a_4}{4}\,\frac{t_1\? t_3 - t_2\? t_4}{(z_1 - z_2)\? (z_2 - z_3)}\,,
		\qquad
		f_2 = \frac{a_4}{4}\,\frac{t_1 + t_3 - t_2 - t_4}{(z_1 - z_2)\? (z_2 - z_3)}\,,
		\CR
		n_3 = &\,  \frac{a_4}{4}\,\frac{t_1\? t_4 - t_2\? t_3}{(z_2 - z_3)\? (z_3 - z_1)}\,,
		\qquad
		f_3 =  \frac{a_4}{4}\,\frac{t_1 + t_4 - t_2 - t_3}{(z_2 - z_3)\? (z_3 - z_1)}\,.
		\end{align}
		Note that \eqref{eq:xyTri} is a change of variables for $x(\rho,z)$ and $y(\rho,z)$  where $r_i(\rho,z) = \sqrt{ (z-z_i)^2 + \rho^2}$ are not all independent, but satisfy \begin{equation}
		(z_3 - z_2)\,r_1^2 + (z_1 - z_3)\,r_2^2 + (z_2 - z_1)\,r_3^2 + (z_3 - z_2)\,(z_1 - z_3)\,(z_2 - z_1) =0\,.
		\end{equation}
		The coordinates $x$ and $y$ are determined by inverting \eqref{eq:Weyl-coord}, and the solution is unique up to the choice of branch for the roots of $\sqrt{r_i^{\; 2}} = \pm r_i$. The eight choices of branch correspond to the eight rectangles defined such that $x$ and $y$ take values in adjacent intervals $[t_{i+1},t_{i}]$ if $t_i > t_{i+1}$, and $(-\infty , t_i ] \cup [t_{i+1},\infty)$ if $t_i< t_{i+1}$.
		This ensures that $\rho\ge 0$ and $z \in \mathds{R}$. 
		
The line \{$\rho=0$, $z\in (-\infty,\infty)$\} is mapped to the boundary of one such rectangle (depending on the chosen branch) in the $(x,y)$ plane. When the function $z(x,y)$ is restricted to one side of this rectangle, it becomes a monotonic function of either $x$ or $y$, so we have a bijection between the boundary of the $(x,y)$ rectangle and the boundary of the compactified $(\rho,z)$ half-plane. The change of variables \eqref{eq:Weyl-coord} therefore defines a homeomorphism from the rectangle $t_2 \le x \le t_1$, $t_3\le y\le t_2$ (with the point $(x,y) = (t_2,t_2)$ removed) to the half-plane $\rho\ge 0$, $z\in \mathds{R}$, and \eqref{eq:xyTri} is its inverse. 

The points common to adjacent rectangles correspond to either ($\rho = 0, z=z_i$) or the point at infinity of the compactified $(\rho,z)$ half-plane. Each one of these four special points is mapped to four isolated points in the $(x,y)$ plane, so there are 16 special points in the $(x,y)$ plane. For a given $i$, at the four points corresponding to ($\rho = 0, z=z_i$) (\ie $r_i= 0$), pairs of branches related by $\sqrt{r_i^{\; 2}} = \pm r_i$ are connected. At the four points in the $x$-$y$ plane that correspond to the point at infinity of the $(\rho,z)$ half-plane, pairs of branches related by all the $r_i$ changing sign are connected. We conclude that the checkerboard of rectangles of adjacent intervals for $(x,y)$ in $\mathds{R}^2$ is  homeomorphically mapped to eight copies of the half-plane $\rho \ge 0$, $z\in \mathds{R}$ (plus the point at infinity).

		We also provide the expansion of the coordinates $x$, $y$, to the first nontrivial order in the spherical coordinates \eqref{eq:Weyl-to-r} around each special point,
		\begin{align}\label{eq:asym-expansions}
		\text{Asympt. infinity: }&\,\quad    
		x = t_2- a_4\, (t_1 - t_2) (t_2 - t_3) (t_2 - t_4)\,\frac{\cos\theta+1}{4\,r} + {\cal O}\Scal{ \frac{1}{r^2}}\,, 
		\CR   &\,\quad    
		y = t_2- a_4\, (t_1 - t_2) (t_2 - t_3) (t_2 - t_4)\,\frac{\cos\theta-1}{4\,r} + {\cal O}\Scal{ \frac{1}{r^2}}\,,  
		\CR
		\text{Centre 1: }&\,\quad  
		x = t_1 + \frac{t_1 - t_2}{a_4\, (t_2 - t_3) (t_2 - t_4)}\,(\cos\theta_1+1)\,r_1 + {\cal O}\scal{r_1^2}\,, 
		\nonumber \rule{0pt}{5ex}\\
		&\,\quad     
		y = t_2 - \frac{t_1 - t_2}{a_4\, (t_1 - t_3) (t_1 - t_4)}\,(\cos\theta_1-1)\,r_1 + {\cal O}\scal{r_1^2}\,,  
		\CR
		\text{Centre 2: }&\,\quad  
		x = t_1 - \frac{t_1 - t_3}{a_4\, (t_2 - t_3) (t_3 - t_4)}\,(\cos\theta_2 - 1)\,r_2 + {\cal O}\scal{r_2^2}\,, 
		\nonumber \rule{0pt}{5ex}\\
		&\,\quad     
		y = t_3 - \frac{t_1 - t_3}{a_4\, (t_1 - t_2) (t_1 - t_4)}\,(\cos\theta_2 + 1)\,r_2 + {\cal O}\scal{r_2^2}\,,
		\CR
		\text{Centre 3: }&\,\quad  
		x = t_2 - \frac{t_2 - t_3}{a_4\, (t_1 - t_3) (t_3 - t_4)}\,(\cos\theta_3 + 1)\,r_3 + {\cal O}\scal{r_3^2}\,, 
		\nonumber \rule{0pt}{5ex}\\
		&\,\quad     
		y = t_3 + \frac{t_2 - t_3}{a_4\, (t_1 - t_2) (t_2 - t_4)}\,(\cos\theta_3 - 1)\,r_3 + {\cal O}\scal{r_3^2}\,.
		\end{align}
		We note that one may easily generate the relevant expansion to any desired order using \eqref{eq:Weyl-coord} and \eqref{eq:Weyl-to-r}. We make both explicit and implicit use of these expansions at various points in the main text.
		
		\newpage
		\section{Vector fields}
		\label{app:vec-fields}
		
		In this appendix we present the vector fields supported by the solution given in Section \ref{sec:solution}. These are obtained straightforwardly by inserting the expressions for the functions $V$, $\Vb$, $K_I$, $L^I$ and $M$ in the corresponding formulae given in \cite{Bena:2016dbw}. The result is organised in terms of a basis of eight independent vector fields; all relevant vector fields may be written as appropriate linear combinations in this basis. 
		
		This basis of vector fields, which due to axisymmetry have only a single component, along $\varphi$, takes the form
		\begin{align}\label{eq:vec-basis}
		W^0_x =&\, \frac{X}{2\? (y^2 X + x^2 Y)}\,d\varphi\,, \qquad W^0_y = \frac{Y}{2\? (y^2 X + x^2 Y)}\,d\varphi\,,
		\CR
		W^1_x =&\, \frac{y\? X}{2\? (y^2 X + x^2 Y)}\,d\varphi\,, \qquad W^1_y = \frac{x\? Y}{2\? (y^2 X + x^2 Y)}\,d\varphi\,,
		\CR
		W^2=&\,  \frac{y^2 X - x^2 Y}{2\? (y^2 X + x^2 Y)}\,d\varphi\,,
		\CR
		W^3_x =&\, \frac{y^3\? X}{2\? (y^2 X + x^2 Y)}\,d\varphi\,, \qquad W^3_y = \frac{x^3\? Y}{2\? (y^2 X + x^2 Y)}\,d\varphi\,,
		\CR
		W^4=&\,  \frac{y^4 X - x^4 Y}{2\? (y^2 X + x^2 Y)}\,d\varphi\,.
		\end{align}
		There are three distinguished vector fields corresponding to the following conserved currents in terms of the functions $V$, $\Vb$ that describe the gravitational instanton,
		\begin{equation}
		\star d\Omega = \frac{\Vb\,dV - V\,d\Vb}{(1+V\Vb)^2} \,, \qquad 
		\star d {\cal W} = \frac{d V + V^2\,d\Vb}{(1+V\Vb)^2} \,, \qquad 
		\star d \bar{\cal W}  = \frac{d\Vb + \Vb^2\,dV}{(1+V\Vb)^2}\,,
		\end{equation}
		which are given in the above basis by
		\begin{align}\label{eq:grav-inst-vec}
		\Omega =&\, 
		\frac1{2\,\delta}\? a_1\? \left( (\delta - 1)\? W^1_x + (\delta + 1)\? W^1_y \right) + \frac1{2\,\delta}\? a_3\? \left( (\delta + 1)\? W^3_x + (\delta - 1)\? W^3_y \right) \,, 
		\CR
		{\cal W}  =&\, 
		-\delta\?a_0\?\left(  W^0_x - W^0_y \right) - \frac1{2\,\delta}\? a_1\? \left( (\delta ^2 + 2\?\delta - 1)\? W^1_x - (\delta ^2 - 2\?\delta - 1)\? W^1_y \right) 
		\CR
		&\,  
		+ \frac1{2\,\delta}\? a_3\? \left( (\delta ^2 - 2\?\delta - 1)\? W^3_x +(\delta ^2 + 2\?\delta - 1)\? W^3_y \right) + \delta\?a_4\,W^4\,,
		\CR
		\bar{\cal W}  =&\, 
		\frac1{2\,\delta}\? a_1\? \left(  W^1_x - W^1_y \right) - \frac1{2\,\delta}\? a_3\? \left(  W^3_x - W^3_y \right) \,.
		\end{align}
		From these expressions, one can see that the basis vector fields $W^3_x$, $W^3_y$ and $W^4$ can be replaced by linear combinations of $\Omega$, ${\cal W}$ and $\bar{\cal W}$ with the remaining five basis elements in \eqref{eq:vec-basis}. We will indeed use these combinations below, in order to exhibit the particular embedding of the currents describing the gravitational instanton within the supergravity solution.
		
		With these definitions, we now proceed to display the various vector fields arising from the solution in Section \ref{sec:solution}, in terms of the basis elements in \eqref{eq:vec-basis}, starting from the $v_I$, which read
		\begin{equation}
		v_I{} = k_I \? \Omega + h_I \? {\cal W} - \frac{a_1}{2\? \delta}\? m_I\? \left( (\delta - 1)\? W^1_x + (\delta + 1)\? W^1_y\right)\,.
		\end{equation}
		Similarly, the vector fields $w^I$ read
		\begin{align}
		w^I =&\, \frac12\?C^{IJK}h_J\?k_K\? \Omega + \frac14\?C^{IJK}h_J\?h_K\? {\cal W} 
		+ \frac14\?\left( C^{IJK}k_J\?k_K  + \frac{a_1^2\? a_4}{a_0\? a_3^2 - a_1^2\? a_4}\?C^{IJK}m_J\?m_K\right)\?\bar{\cal W}
		\CR
		&\,
		- \frac{a_1}{4}\? C^{IJK}h_J\?m_K\? \left(  W^1_x + W^1_y\right)
		-\frac{1}{8\?\delta}\?\frac{a_1^2\? a_3^2}{a_0\? a_3^2 - a_1^2\? a_4}\? C^{IJK}m_J\?m_K \? W^2 
		\CR
		&\,
		-\frac{a_1}{4\?\delta}\?\left( C^{IJK}(k_J - h_J)\?k_K  + \frac{a_1^2\? a_4}{a_0\? a_3^2 - a_1^2\? a_4}\?C^{IJK}m_J\?m_K\right)\?\left(  W^1_x - W^1_y\right)
		\,.
		\end{align}
		Finally, we provide the expressions for the vector field $w^0$,
		\small
		\begin{align}
		w^0 =&\, (q_0 + l_0 - \tfrac12\? l^I\?k_I) (\Omega + \bar{\cal W})
		+ (q_0 + \tfrac18\? C^{IJK}\?h_I\?h_J\?k_K) \?\Omega  + (q_0 + \tfrac14\? h_1\? h_2\? h_3) \?{\cal W}
		\CR
		&\,
		+\frac14\?\left( \tfrac12\?C^{IJK}\?h_I\?k_J\?k_K - k_1\? k_2\? k_3 \right) \? \bar{\cal W}
		+\frac{a_0\? a_1^3\? a_3^3}{16\? \delta\? (a_0\? a_3^2 - a_1^2\? a_4)^2}\? m_1\? m_2\? m_3\? (W^0_x - W^0_y)
		\CR
		&\,
		+\frac14\?a_1\? \left(l^I\?m_I -\tfrac14\? C^{IJK}\?h_I\?h_J\?m_K \right)\? ( W^1_x + W^1_y )
		\CR
		&\,
		-\frac14\? \frac{a_1^2\? a_4}{a_0\? a_3^2 - a_1^2 a_4}\? 
		\left(\tfrac12\?C^{IJK}\?m_I\?m_J\?(k_K-h_K) + 
		\frac{2\? a_1^2\? a_4}{a_0\? a_3^2 - a_1^2\? a_4}\? m_1\? m_2\? m_3 \right)\? 
		\left( \bar{\cal W} - \frac{a_1}{\delta}\, (W^1_x - W^1_y) \right)
		\CR
		&\,
		+\frac{a_1}{4\? \delta}\? \left( \tfrac14\? C^{IJK}\?m_I\?(k_J-h_J)\?(k_K-h_K) + 
		\frac{a_1^2\?a_4}{2\? (a_0\? a_3^2 - a_1^2\? a_4)}\? m_1\? m_2\? m_3 \right)\? (W^1_x - W^1_y)
		\CR
		&\,
		+\frac1{16\? \delta}\? \frac{ a_1^4\? a_3^2}{(a_0\? a_3^2 - a_1^2\? a_4)^2}\? m_1\? m_2\? m_3\?
		\left(\frac12\? a_3\? (W^1_x - W^1_y) + 3\? a_4\? W^2 \right)
		\CR
		&\,
		+\frac1{32\? \delta}\? \frac{a_1^2\? a_3^2}{a_0\? a_3^2 - a_1^2 a_4} \?
		\?C^{IJK}\?m_I\?m_J\?(k_K-h_K)\? W^2
		\,.
		\end{align}
		\normalsize
		and for the vector field $\omega$,
		\small
		\begin{align}
		\omega =&\, -(q_0 + l_0 - \tfrac12\? l^I\?k_I) \?\Omega -  q_0\? {\cal W}
		+ \frac14\? k_1\? k_2\? k_3 \?\bar{\cal W}
		\CR
		&\,+ \frac14\? \frac{a_1^2\? a_4}{a_0\? a_3^2 - a_1^2 a_4}\? 
		\left(\tfrac12\?C^{IJK}\?m_I\?m_J\?(k_K-h_K) + 
		\frac{2\? a_1^2\? a_4}{a_0\? a_3^2 - a_1^2\? a_4}\? m_1\? m_2\? m_3 \right)\? 
		\left( \bar{\cal W} - \frac{a_1}{\delta}\, (W^1_x - W^1_y) \right)
		\CR
		&\,
		-\frac{a_0\? a_1^3\? a_3^3}{16\? \delta^2\? (a_0\? a_3^2 - a_1^2\? a_4)^2}\? m_1\? m_2\? m_3\? 
		\left( (\delta-1)\?W^0_x - (\delta+1)\?W^0_y \right)
		\CR
		&\,
		-\frac14\?a_1\? \left(l^I\?m_I - \frac{a_1^3\? a_3^3}{8\? \delta^2\? (a_0\? a_3^2 - a_1^2\? a_4)^2}\? m_1\? m_2\? m_3 \right)\? ( W^1_x + W^1_y )
		\CR
		&\,
		+\frac{a_1}{4\? \delta}\? \left( l^I\?m_I - \tfrac14\? C^{IJK}\?k_I\?k_J\?m_K - 
		\frac{a_1^2\?a_4}{2\? (a_0\? a_3^2 - a_1^2\? a_4)}\? m_1\? m_2\? m_3 \right)\? (W^1_x - W^1_y)
		\CR
		&\,
		-\frac1{16\? \delta}\? \frac{ a_1^4\? a_3^2}{(a_0\? a_3^2 - a_1^2\? a_4)^2}\? m_1\? m_2\? m_3\?
		\left(\frac12\? a_3\? (W^1_x - W^1_y) + 3\? a_4\? W^2 \right)
		\CR
		&\,
		-\frac1{32\? \delta}\? \frac{a_1^2\? a_3^2}{a_0\? a_3^2 - a_1^2 a_4} \?
		\?C^{IJK}\?m_I\?m_J\?k_K \? W^2
		\,.
		\end{align}
		\normalsize

	\end{appendix}
	
	\newpage
	\parskip=0pt
	\baselineskip=14pt
	
	\bibliography{PaperG}
	\bibliographystyle{utphys}

\end{document}